\documentclass[pra,twocolumn,floatfix,superscriptaddress,longtable,nofootinbib,notitlepage,10pt]{revtex4-2}
\usepackage{amsmath}
\usepackage{lipsum} 
\usepackage{verbatim}
\usepackage{epsfig}
\usepackage{subfigure}
\usepackage{graphicx}
\usepackage{enumitem}
\usepackage{amsfonts}
\usepackage[final]{showlabels}
\usepackage{enumitem}
\usepackage[figuresright]{rotating}
\usepackage{amssymb}
\usepackage{amsmath}
\usepackage{psfrag}
\usepackage{mathtools}
\usepackage[export]{adjustbox}
\usepackage{bm}
\usepackage[colorlinks,linkcolor=blue,anchorcolor=blue,citecolor=blue,urlcolor=blue]{hyperref}
\usepackage[version=4]{mhchem}
\usepackage[svgnames]{xcolor}

\def\be{\begin{equation}} \def\ee{\end{equation}}
\def\bea{\begin{eqnarray}} \def\eea{\end{eqnarray}}

\def\bpm{\begin{pmatrix}} \def\epm{\end{pmatrix}}

\makeatletter
\newcommand*{\balancecolsandclearpage}{%
	\close@column@grid
	\clearpage
}
\makeatother

\begin{document}
	\title{Quantum Darwinism-encoding transitions on expanding trees}
	
	\author{Beno\^it Fert\'e}
	\affiliation{Laboratoire de Physique de l'\'Ecole normale sup\'erieure, ENS, Universit\'e PSL, CNRS, Sorbonne Universit\'e, Universit\'e Paris Cit\'e, F-75005 Paris, France}
	
	\affiliation{Universit\'e Paris-Saclay, CNRS, LPTMS, 91405, Orsay, France}
	
	\author{Xiangyu Cao}
	\affiliation{Laboratoire de Physique de l'\'Ecole normale sup\'erieure, ENS, Universit\'e PSL, CNRS, Sorbonne Universit\'e, Universit\'e Paris Cit\'e, F-75005 Paris, France}
	
	\begin{abstract}
		Quantum Darwinism (QD) proposes that classical objectivity emerges from the broadcast of information about a microscopic degree of freedom into multiple fractions of a many-body environment. Such a broadcast of information is in sharp contrast with its scrambling under strong interaction. It was recently shown that quantum dynamics interpolating between broadcasting and scrambling may display sharp phase transitions of information propagation, named QD-encoding transitions. Here, we initiate their systematic study in generic, non-Clifford settings. First, in a general theoretical setup where the information propagation is modeled as an isometry, whose input qudit is entangled with a reference, we propose a probe of the transitions --- the distribution of the density matrix of the reference after measuring an environment fraction. This probe measures the classical correlation between the fraction and the injected information. We then apply the framework to two similar models defined by a tensor network on an expanding tree, modeling a noisy apparatus that attempts to broadcast the $z$ component of a spin-half. We derive an exact recursion relation of the density matrix distribution, which we analyze analytically and numerically. As a result we find three phases: QD, intermediate and encoding, and two continuous transitions. The encoding-intermediate transition describes the establishment of nonzero correlation between the reference and a small environment fraction, and can be probed by a ``coarse-grained'' measure of the total spin-$z$ of the fraction, which becomes non-Gaussian and symmetry breaking in the intermediate space. The QD-intermediate transition is about whether the correlation is perfect. It must be probed by fined-grained measures, and corresponds to a more subtle symmetry breaking in the replica space. 
	\end{abstract}
	
	\date{\today}
	\maketitle
	
	\section{Introduction}
	Why does the macroscopic world surrounding us appear classical, although it obeys the laws of quantum mechanics, to the best of our knowledge? This basic question, raised since the birth of quantum mechanics, remains unsettled. An important step was taken by the decoherence theory~\cite{zurek-decoherence,schloss-deco}, which points out that any quantum system is inevitably in contact with some environment, and becomes entangled to it. As a result, a coherent superposition rapidly evolves into a probabilistic mixture of classical ``pointer states''. This begs in turn the question of which pointer states can survive decoherence. 
	
	To address this question, a promising approach, much explored in the past decades~\cite{ollivier-poulin,blume-kohout-zurek,zurek-QD,Ryan-onion,campbell,collision-model,paternostro18-darwin,le-olaya-pra,le-prl19,unden19-darwin-exp,zurek-review,Korbicz-rev}, consists in examining how information about a quantum system propagates in its environment, whose complex many-body structure must be taken into account. This quantum information-theoretical analysis of the system-environment universe yields an important insight: \textit{some} information about the system --- that corresponding to pointer states --- is duplicated and broadcast into the environment. Thus, many observers, each having access to a small fraction of the environment, are able to retrieve the information and agree on it: the information becomes objective.

	An illustrative example of an objective fact is a measurement result. Suppose that a qubit (the system) is measured in the computational basis by a macroscopic apparatus in a laboratory (the environment). From a super-observer's point of view in a Wigner's friend thought experiment (see e.g.~\cite{bruckner,wiseman} for recent advances), it becomes entangled with its environment, in a way that the whole laboratory is approximately in a Greenberger-Horne-Zeilinger~\cite{greenberger1989} state: 
	\begin{equation*}
		\vert \Psi \rangle_{\text{lab}} \approx \frac1{\sqrt{2}} \left( | 0 \rangle_{\text{qubit}} | 0 \dots 0 \rangle_{\text{env.}} +  | 1 \rangle_{\text{qubit}} | 1 \dots 1 \rangle_{\text{env.}}  \right) \,.
	\end{equation*}
	In this simple example, the pointer states are $|0\rangle$ and $| 1 \rangle$. Any fraction of the environment, even a single bit, is perfectly correlated to the system. Measuring any environment bit (in the computational basis) will disentangle the system bit, and make it collapse into one of the pointer states. In this sense, the information of whether the qubit is in $|0 \rangle$ or $|1\rangle$ has become objective and retrievable in many small fractions of the environment. By contrast, one would need to measure the whole environment to make the system bit collapse to a non-pointer state, for example, $(|0\rangle + |1\rangle) / \sqrt{2}.$ 
	
	Now, a \textit{typical} random state in the system-environment Hilbert space has a completely different structure of correlation. It is well-known that, although the system is maximally entangled to the environment, a small fraction of the environment (smaller than half of the latter) is \textit{uncorrelated} with the system~\cite{page,nielsen_chuang_2010,preskilllecture,schumacher,schumacher-QEC}. In other words, all information about the system is ``encoded'', and inaccessible for all practical purposes. The encoding  of information is also a property of \textit{generic} many-body unitary evolution, that is, generated by a non-integrable Hamiltonian~\cite{deutsch,srednicki,rigol-review}. Information injected by a local perturbation, while conserved by unitary, becomes more and more non-local and inaccessible. Such information encoding (often known as ``scrambling'') is believed to underlie the emergence of irreversible phenomena such as thermalization and hydrodynamic relaxation~\cite{Sekino_2008,haydenpreskill,shenkerstanford,swingle-review}. 
	
	We have thus two diametrically opposite patterns of quantum information spreading. They are both --- or at least, expected to be ---  ``generic'', but in different contexts. Encoding is generic when the ``universe'' consists of an isolated strongly interacting quantum system~\cite{RevModPhys.91.021001}; such a ``universe'' may be realized with controlled experiments~\cite{ultracold,trappedion,DOHERTY20131}, strongly correlated materials at low temperatures~\cite{hartnoll2022colloquium}, and (simulated) black holes~\cite{Sekino_2008,qginthelab}. Meanwhile, classical objectivity arises generally in ``universes'' that are more familiar to us, such as a laboratory apparatus measuring a quantum spin. Usually we do not model such a universe as a many-body quantum system; instead, we trace out most of the ``bath'' degrees of freedom and focus on the resulting dissipative dynamics~\cite{breuer-book}. Nevertheless, in principle, both classical objectivity and encoding behaviors emerge in some many-body quantum dynamics. Hence it is natural to ask whether they can be understood as ``phases of (quantum) information'', and identify phase transitions between them. More broadly speaking, these questions can be thought of part of the  more ambitious goal of classifying ``phases of information'', as an extension to the classification of phases of matter in equilibrium statistical mechanics.

	While earlier works on Quantum Darwinism~\cite{riedel2012,campbell,duruisseau} already pointed out that an environment can exhibit encoding or QD behaviors upon adjusting some parameter, or in different time regimes, the task of addressing them as phases of information and studying transitions between them was first undertaken recently in Ref.~\cite{FC}. This work proposed a toy model where one bit of quantum information propagates in a structured environment, which is modeled as a random \textit{Clifford} unitary circuit on an expanding tree, depending on one parameter. By tuning the latter, the environment can be either in an ``encoding'' phase where the injected information is inaccessible in any environment fraction (unless it is larger than half of the environment), or a ``Quantum Darwinism'' (QD) phase where it is accessible in arbitrarily small environment fractions. The two phases are separated by a stochastic-mixed phase where a random instance of the ``universe'' can be either encoding or QD with nonzero probability. Two continuous phase transitions emerge at the QD-mixed and mixed-encoding boundaries, respectively. 
	
	The choice of Clifford circuit was mainly motivated by solvability, but also limited the generality of the findings. Indeed, in Clifford stabilizer states, quantum correlation is ``quantized''~\cite{gottesman1998heisenberg,aaronson-gottesman}: for example, two qubits can only be completely uncorrelated, maximally entangled, or classically correlated in the Pauli $Z$, $X$ or $Y$, direction. Of course, in general, a continuum of other possibilities can exist. Thus, the quantity used to distinguish the different phases in the Clifford model does not apply beyond Clifford. Also, the stochastic-mixed phase is most probably specific to Clifford models, where entanglement entropy and mutual information can only be an integer times $\ln 2$.

	\begin{figure}
		\centering
		\includegraphics[scale=1.]{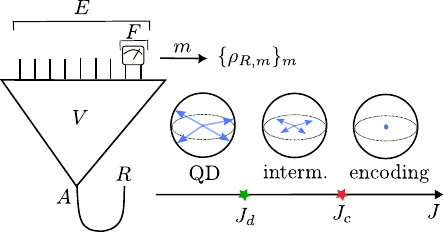}
		\caption{General theoretical setup for studying information propagation in an environment. $V$ is an isometry from $A$ (injected qudit) to $E$ (output environment). $A$ is initially entangled with a reference $R$. To probe the classical correlation between a fraction $F \subset E$ and $R$, we perform a measurement on $F$, and consider the the post-measurement density matrix of $R$. By definition, in the Quantum Darwinism (QD) phase, $\rho_{R,m}$ is almost surely a pure state; in the encoding phase, $\rho_{R,m} $ is almost surely maximally mixed. (The density matrices are represented using the Bloch sphere.) In the tree models studied in this work, the two phases appear at small and large values of a ``scrambling parameter'' $J$ (that controls $V$). The two phases are separated by an intermediate phase and two transitions. See Figs.~\ref{fig:Ps} and \ref{fig:basic} below.}
		\label{fig:gen}
	\end{figure}
	In this paper, which is a follow-up on Ref.~\cite{FC}, we initiate a systematic study of QD-encoding phase transitions beyond Clifford models. First, in Section~\ref{sec:gen}, we propose a probe of QD-encoding transitions that applies to a general class of models of information propagation in a many-body environment (see Fig.~\ref{fig:gen}). We adopt a widely used theoretical technique of keeping a quantum copy of the injected information in a reference qudit~\cite{haydenpreskill,gullans-huse-prl}. Then, the probe measures what we can learn about the reference (and thus the injected information) by measuring a fraction of the environment. In other words, the probe concerns the ``classical'' part of the system-environment correlation~\cite{Henderson_2001,discord-zurek,girolami}, or the Holevo bound~\cite{Holevo}. This informtion-theoretical notion is directly related to the ensemble of random post-measurement density matrices of the reference, where the randomness comes from the Born's rule (and eventually the randomness of the model itself). Thus, we define the phases of information in terms of the random density matrix distribution in the in the thermodynamic limit: the QD, encoding, and intermediate phases correspond to a distribution of pure, maximally mixed, and partially mixed states, respectively.  While the density matrix distribution can only be a sum of finite number of delta peaks in a Clifford model, it has a continuum support and nontrivial form in general.
	
	The second contribution of this work is a detailed analysis of two similar non-Clifford expanding-tree models (Section~\ref{sec:trees}); one of them is \textit{deterministic}. They can be viewed as idealized models of an apparatus/environment attempting to broadcast the $z$-component of the input spin-half (qubit). Although intuitively similar to their Clifford cousin~\cite{FC}, the generic models avoid the latter's artifacts, and are also more technically involved. To analyze them, we derive an exact recursion relation satisfied by the density matrix distribution. The result here applies to a large class of hierarchical models. The recursion relations can be viewed as an analog of the ``traveling wave equation'' routinely used to analyze branching and tree models, see for example~\cite{derrida-spohn,brunet,feng2022measurement,Ferte_2023}. Here, in addition to being nonlinear, our ``traveling wave equation'' is also non-local, making its solution a formidable challenge. 
	
	Nevertheless, combining analytical and numerical techniques, we established the phase diagram of both models. They turn out to resemble that of the random Clifford model: as a function of a ``scrambling'' parameter $J \in (0,1)$, both models display three phases --- QD ($J < J_d$), intermediate ($J_d < J < J_c$) and encoding ($J > J_c$) --- separately by two critical points $J_d$ and $J_c$ (see Fig.~\ref{fig:gen}). 
	
	The encoding phase is, similarly to that in the Clifford model, characterized by the absence of correlation between any environment fraction and the reference (unless the fraction is larger than half the environment). Meanwhile, the QD and intermediate phases are qualitatively distinct from the Clifford case. In the intermediate phase, which is no longer a stochastic mixture, measuring an environment fraction partially disentangles the reference, revealing \textit{some} of the injected information: its amount is independent of the fraction's relative size in the thermodynamic limit. In the QD phase, this amount becomes the maximal value, one qubit: the reference is completely disentangled upon measuring the fraction. Yet, its posterior polarization direction, randomly distributed according to Born's rule, is \text{close}, but not exactly equal, to $| \pm_z \rangle$ (unless $J = 0$). This noisy selection of the pointer state has the following consequence: if the input qubit is not entangled with the reference, but prepared by Alice to be either $| +_z \rangle$ or $| -_z \rangle$, then Bob cannot infer Alice's choice with perfect certainty. 
	
	The nature of the phases of information dictates, to a large extent, that of the critical points. The encoding-intermediate transition is one between zero and nonzero information retrieval. We are able to analytically locate the critical point and characterize its critical properties (they are of simple mean-field nature). Moreover, we point out that this transition is easy to probe. It suffices to measure a ``coarse-grained'' observable of a small fraction {of the environment}, its total spin (the $z$-component), $\mathcal{M}$. In the encoding phase $\mathcal{M}$ has a Gaussian statistics and is uncorrelated with the reference; in the intermediate phase, $\mathcal{M}$ is correlated with the reference, and has a non-Gaussian distribution with two peaks. In this regard the encoding-intermediate transition is of a conventional kind, associated with the breaking/restoration of a $\mathbb{Z}_2$ symmetry.
	
	The QD-intermediate transition, between imperfect and perfect correlation, is more subtle. For instance, we will show that it cannot be probed with a ``coarse-grained'' measurement. To understand heuristically the nature of this transition, we may observe that it is a purification transition: observing the environment fraction completely (partially, respectively) disentangles the reference in the QD (intermediate, respectively) phases. Hence, we may compare the QD-intermediate transition to the measurement-induced transitions~\cite{skinner19,li-fisher,MIPTrev}, which are also characterized by purification~\cite{gullans-huse-prx,gullans-huse-prl}. Such transitions are known to be associated with a more abstract symmetry breaking, in the replica space~\cite{bao-altman,nahum21,vasseur-ludwig,nahum2023renormalization}. In this work, we will \textit{not} use the replica trick. Instead, we provide a direct characterization of the replica-symmetry breaking in terms of the density matrix distribution and its ``equation of motion'', given by recursion relations (in this sense, our approach is reminiscent of the cavity method in spin glass theory~\cite{spinglass}). The equation of motion always has a QD (replica-symmetric) solution, but it becomes unstable in the intermediate (replica-symmetric breaking) phase. This idea allowed us to numerically locate the critical point with good precision despite the pronounced finite-size effects and the absence of exact solution. While numerical data are compatible with standard mean-field critical behaviors, an analytical understanding of the generic QD-intermediate transition remains an open question. 
	
	\tableofcontents
	
	\section{General setup and observables}\label{sec:gen}
	In this Section, we first propose a general theoretical setup for studying phases of information propagation in structured environments (Section~\ref{sec:setup}). Then we shall define the phases of information:  QD, encoding and intermediate, in terms of the Holevo bound (Section~\ref{sec:probedef}). Section~\ref{sec:RDME} introduces the notion of random density matrix ensemble. It is closely related to the Holevo bound, and provides an equivalent definition of the information phases. The above Sections are essential to understand the rest of the paper. Meanwhile, readers may skip Section~\ref{sec:mutual}, which reviews the relation to mutual information, as well as Section~\ref{sec:Clifford-gen}, which is about the special case of Clifford models~\cite{FC}.
	{ 
		\subsection{Setup}\label{sec:setup}
		We consider the propagation of one qudit of quantum information injected into a many-body environment. Upon enlarging the environment, we may assume that the process is isolated, and thus described by an isometry from the Hilbert space of the injected qudit $A$, of dimension $q$, to that of the environment $E$ by the end of the process:
		\begin{equation}
			V: \mathbb{C}^q \simeq \mathcal{H}_A  \to \mathcal{H}_{E} \,,\, V^\dagger V = \mathbf{1} \,.
		\end{equation}
		Using standard tensor network notation, we may represent $V$ by an triangle, and the isometry identity as follows: 
		\begin{equation}
			V = \includegraphics[scale=1,valign=c]{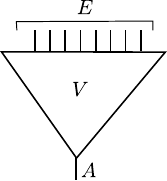} \, \;,\; \includegraphics[scale=1,valign=c]{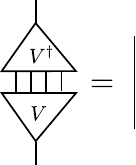} \,.
		\end{equation}
		Here the outgoing environment $E$ will have many degrees of freedom, represented by several legs. Note that the injected qudit $A$ may not be part of $E$. This is the case in a destructive measurement apparatus, for instance a photon detector, which destroys the incident photon.
		
		It is sometimes useful~\cite{FC} to view the isometry $V$ as being obtained by a unitary map $U$ which maps $A$ and some incoming environment degrees of freedom to the outgoing environment $E$:
		\begin{equation}\label{eq:VfromU}
			\includegraphics[scale=1,valign=c]{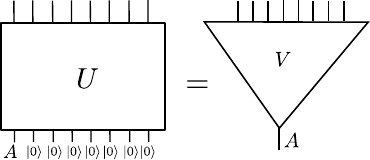} \,,
		\end{equation}
		upon contracting with the incoming environment state, which we assume to be a pure factorized one. However, in what follows, we shall focus on the isometry $V$ and the term ``environment'' will always refer to the final one, $E$.  
		
		In order to study the correlation between (a fraction of) $E$ and the $A$, it is convenient to introduce the Choi-Jamiolkowski (CJ) state of the isometry $V$, denoted by $\Psi_V$~\cite{CHOI1975285,JAMIOLKOWSKI1972275}. Recall that this is obtained by entangling initially the injected qudit $A$ with a reference ($R$), and applying $V$ on $A$, leaving $R$ intact:
		\begin{equation}
			| \Psi_V \rangle = (\mathbf{1}_R \otimes V_{A}) | I \rangle_{RA} \,,\,   | I \rangle_{RA} =  \frac1{\sqrt{q}}\sum_{i=0}^{q-1}  \vert i \rangle_R  |  i \rangle_A \,.
		\end{equation}
		A graphical representation of the CJ state is as follows:
		\begin{equation}\label{eq:PhiV-graph}
			| \Psi_V \rangle =  \includegraphics[scale=.9,valign=c]{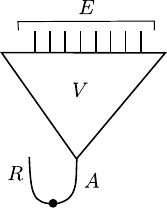}
		\end{equation}
		where a black dot stands for a $1/\sqrt{q}$ factor. By isometry, $R$ remains maximally entangled to $E$, so that the reduced density matrix of $R$ is maximally mixed: 
		\begin{equation}
			\rho_R = \frac1{q} \mathbf{1} = \; \includegraphics[scale=1,valign=c]{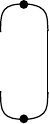}\label{eq:rhoR-prior}
		\end{equation}
		The information-theoretical meaning of the CJ state is the following: The reference qudit keeps a quantum record of the injected information available after the propagation process destroys $A$. Thus, we can address the correlation between the environment and $R$ (the injected information) within the CJ state $\Psi_V$, as we see below. 
		
		We remark that this way of characterizing correlation is routinely used in other contexts, such as in the black hole information problem, where one is interested in the correlation between the Hawking radiation and the information carried by an in-falling object~\cite{haydenpreskill}. The ``reference bit'' method also proved useful in characterizing measurement-induced phase transitions~\cite{bao-altman,gullans-huse-prl}.
		
		\subsection{Phases of information}\label{sec:probedef}
		Following the approach of quantum Darwinism and similar approaches to emergent classicality, we shall consider what information on $R$ can be revealed from performing some measurement on a \textit{fraction} (subsystem) of the environment, $F \subset E$. Suppose that we a measurement outcome $m$. The reduced density matrix of the reference qudit is then updated from the the maximally mixed one $\rho_R$ \eqref{eq:rhoR-prior} to $\rho_{m}$:
		\begin{equation}
			\includegraphics[scale=1,valign=c]{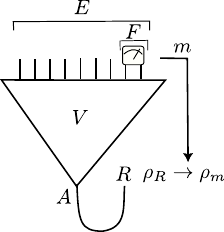} \,.
		\end{equation}
		
		A well-known quantity in quantum information under the names of Holevo bound~\cite{Holevo}, asymmetric mutual information, and classical correlation~\cite{Henderson_2001,discord-zurek}, measures the expected amount of information revealed by the measurement. By definition, it is equal to the von Neumann entropy decrease, averaged over the measurement results:
		\begin{equation}\label{eq:JFRdef}
			\chi(F;R) =  S(\rho_R) - \sum_m p_m S(\rho_m)  \,. 
		\end{equation}
		Here, $p_m$ is the probability of the outcome $m$, and $S(\rho) = - \mathrm{Tr}[\rho \ln \rho]$ is the von Neumann entropy. Since $S(\rho_R) = \ln q$ \eqref{eq:rhoR-prior} and $0 \le S(\rho_m) \le q$ for any $m$, 
		\begin{equation}
			0 \le \chi(F;R) \le \ln q  \,.
		\end{equation}
		We shall therefore use $\chi(F;R) $ to define three phases of information retrieval:
		\begin{itemize}
			\item A \textbf{Quantum Darwinism} (QD) phase is one in which 
			\begin{equation}
				\chi(F;R) \to \ln q 
			\end{equation}
			in some thermodynamic limit (same below). In the QD phase, measuring the environment fraction reveals all of the injected information~\cite{girolami,duruisseau}. 
			\item An \textbf{encoding} phase is one in which 
			\begin{equation}
				\chi(F;R) \to 0 \,.   
			\end{equation}
			In the encoding phase, measuring the environment fraction fails to reveal any injected information. 
			\item An \textbf{intermediate} phase is one in which none of the above holds. In this phase, measuring the environment reveals partially the injected information.
		\end{itemize}
		The use of the term ``phase'' is justified by the following observation: As a model goes from one phase to another upon tuning some parameter, the thermodynamic limit of $\chi(F;R)$ must depend non-analytically on that parameter (an analytical function that is constant somewhere must be constant everywhere). \textit{Phase transitions} of information retrieval are, by the standard definition, parameter space loci where $\chi(F;R)$ is non-analytical. 
		
		By the above definition, the phase of information retrieval depends on the environment dynamics (described by the isometry $V$), the fraction $F$ and the choice of measurement. As we will see, the phase diagram of a model generally depends on the choice of the measurement. Often, the term Quantum Darwinism is also associated with an independence on the fraction size $F$, especially when $|F| / |E| \to 0$, since classical objectivity requires information to be retrievable in small fractions. We do not include this requirement in the above definition to keep it simple. Instead, we will treat the fraction size as one extra parameter of the phase diagram. Nevertheless, It turns out that in the tree models we shall study (and those of \cite{FC}), the phase of the model does become independent of the relative size $|F| / |E|$ in the thermodynamic limit.

		
		\subsection{Random density matrix ensemble}\label{sec:RDME}
		The Hovelo bound $\chi(F;R)$ quantifies \textit{how much} information is revealed. Taking a step further, we may describe \textit{which} information is likely to be revealed, by the ensemble of post-measurement density matrices $\rho_m$, weighed by their respective outcome probability $p_m$. We find it convenient to rescale the density matrices as
		\begin{equation}\label{eq:rhomQtm}
			\tilde{Q}_m := q \rho_m   \,,
		\end{equation}
		and define the following random matrix ensemble (or distribution)
		\begin{equation}\label{eq:PQdef}
			\mathbb{P}(\tilde{Q}) = \sum_m p_m \delta(\tilde{Q} - \tilde{Q}_m)  \,.
		\end{equation} 
		
		Averages with respect to this distribution will be denoted by $\left< [\dots] \right>$: For any observable $f$ that depends on a matrix $\tilde{Q}$, 
		\begin{equation}\label{eq:observ}
			\left< f(\tilde{Q}) \right> := \sum_m p_m f(\tilde{Q}_m) \,.
		\end{equation}
		By construction, the Holevo bound is such an ensemble average. Indeed, using \eqref{eq:rhoR-prior}, \eqref{eq:chiFR} and \eqref{eq:rhomQtm}, it is not hard to show that:
		\begin{equation} \label{eq:JFR}
			\chi(F; R) = \frac{1}q \left< \mathrm{Tr}[\tilde{Q} \ln \tilde{Q} ] \right>  \,.
		\end{equation}
		
		Since the three phases of information are defined with regard to the extreme values of $\chi(F;R)$, it is not hard to see that they are characterized by the following properties of the random matrix ensemble in the thermodynamic limit: 
		\begin{itemize}
			\item In the QD phase, $\tilde{Q}$ is almost surely of rank one;or equivalently,  $\tilde{Q} / q$ is a pure state). In other words, in the QD phase, measuring $F$ completely disentangles the reference qudit almost surely. A distribution that is supported in the manifold of pure states will be called ``perfectly QD''. 
			\item In the {encoding} phase, $\tilde{Q} = \mathbf{1}$ with probability one. In other words, in the encoding phase, measuring $F$ does not affect the reference qudit. The distribution $\delta(\tilde{Q} - \mathbf{1})$ will be called ``perfectly encoding''.
			\item In the {intermediate} phase, none of the above holds, that is, $\tilde{Q}/q$ is a mixed state but not maximally mixed. An intermediate distribution is one that is neither QD nor encoding.
		\end{itemize}
		Note that the above can be taken as equivalent definition of the phases of information. 
		
		As a consequence, any function $f(\tilde{Q})$ that is extremized by perfectly QD and perfectly encoding ensembles can be used to probe the three phases. Besides the Holevo abound, another example is the  ``purity'' \footnote{Note that this has nothing to do with a two-replica calculation; the ensemble average considered in this work is always the ``physical'' one dictated by Born's rule, or the $n\to 1$ limit in terms of the replica trick}:
		\begin{equation}\label{eq:puritydef}
			r^2  := \frac{ \mathrm{Tr}[\tilde{Q}^2] - q}{q^2 - q}  \,.
		\end{equation} 
		From the equivalent definitions above, it follows that  the averaged purity $\left<r^2 \right>$ tends to $1$ and $0$ in the QD and encoding phase, respectively. For a qubit ($q=2$), the Helovo bound is related to the purity as follows:
		\begin{equation}\label{eq:chiFR}
			\chi(F,R) = \ln 2 + \sum_{s=\pm1} \left< \frac{1+ sr}2 \ln  \frac{1 + sr}2 \right> \, 
		\end{equation} 
		since $(1\pm r)/2$ are the two eigenvalues of the post-measure density matrix. Thus, $\chi$ has the same qualitative behavior as the purity average. More quantitatively, $\chi \sim \left< r^2  \right> / 2 $ when the latter is small; when the latter approaches $1$ with $\left< r^2 \right> = 1 - 2 \delta, \delta \ll1 $, we have a log correction $\chi \to \ln 2 + \delta \ln (\delta) / 2  + O(\delta)$. In what follows, we will prefer to use the simpler $\left< r^2 \right>$ to probe the phases of information.  
		
		We now turn to deriving a few simple general formulas that will be useful for calculating the random matrix ensemble. For this, consider a positive operator valued measurement (POVM) on $F$, specified by a family of positive semi-definite Hermitian operators $\pi_m$, indexed by the measurement outcome, such that the sum equals identity~\cite{peres-review,nielsen_chuang_2010}:
		\begin{equation}\label{eq:sumpim}
			\sum_m \pi_m = \mathbf{1} \,.
		\end{equation}
		Now consider the following operator
		\begin{equation}\label{eq:Qmdef}
			{Q}_m = V^\dagger \pi_m  V = \includegraphics[scale=1,valign=c]{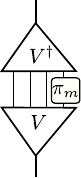}
		\end{equation} 
		which acts on the injected qudit $A$, and which can be viewed as the Heisenberg time evolution of $\pi_m$. Then it follows from the Born's rule that the outcome $m$ is
		\begin{align}  
			p_m = \langle \Psi_V | \pi_m  | \Psi_V \rangle =  
			\langle I_{RA}| Q_m  | I_{RA} \rangle 
			= \frac1q \mathrm{Tr}[{Q}_m]  \label{eq:pmdef}
		\end{align}
		A Graphical representation of the calculation is [see \eqref{eq:PhiV-graph} and \eqref{eq:Qmdef} above]
		\begin{equation*}
			p_m := \includegraphics[scale=1,valign=c]{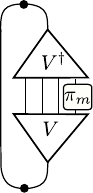} =  \includegraphics[scale=1,valign=c]{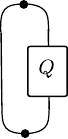} \,.
		\end{equation*}
		Hence, the post-measurement reduced density matrix of the reference qudit is [see also \eqref{eq:rhomQtm} above]: 
		\begin{equation}\label{eq:QtmQm}
			\tilde{Q}_m = q \rho_m  = Q_m / p_m \,.
		\end{equation}
		
		Equations \eqref{eq:pmdef} and \eqref{eq:QtmQm} will allow us to calculate the random matrix ensemble in Section~\ref{sec:trees} below. For now, let us apply them to derive a general property of the random matrix ensemble: the average with $f(\tilde{Q})=\tilde{Q}$ is the identity. Indeed, recalling \eqref{eq:sumpim}, \eqref{eq:Qmdef}, and the isometry of $V$, 
		\begin{align}
			\left< \tilde{Q} \right> =& \sum_m p_m \tilde{Q}_m \nonumber
			= \sum_m Q_m = \sum_m V^\dagger \pi_m V =V^\dagger V  \\ = & \mathbf{1} \,. \label{eq:Qaverageis1}
		\end{align}
		(We leave it to interested readers to draw a graph.) A consequence of this is that the distribution $\mathbb{P}(\tilde{Q})$ can be a single delta peak only in the encoding phase, where the peak is at $\tilde{Q} = \mathbf{1}$. 

	}
	We end this Section with a few remarks. First, it will be useful to consider models defined by a random ensemble of isometries $V$ instead of a single deterministic one. In that case, we adapt the definition \eqref{eq:PQdef} of the density matrix distribution by averaging further over $V$ (denoted by $\mathbb{E}_V$):
	\begin{equation}\label{eq:PRrandom}
		\mathbb{P}(\tilde{Q}) = \mathbb{E}_V \left[ \sum_m  p_m  \delta(\tilde{Q} - \tilde{Q}_m) \right] \,.
	\end{equation}
	Thus, the ensemble average $\left< [\dots] \right>$ is over both the measurement outcome and the realization of $V$. This is a sensible definition because any reasonable ensemble average is a linear functional of the distribution $\mathbb{P}(\tilde{Q})$, even if it typically involves a nonlinear observable in $\tilde{Q}$. In particular, eq.~\eqref{eq:JFR} still holds if we replace the left hand side by the average Holevo bound $\mathbb{E}_V [ \chi(F; R)]$. A similar remark applies to the purity, \eqref{eq:puritydef}. In what follows, we will absorb the average $\mathbb{E}_V $ into the notation $\left< [\dots] \right>$ for brevity.

	Second, by focusing on the random matrix ensemble $\tilde{Q}$, we are ignoring the correlation between the revealed information $\tilde{Q}_m$ and the outcome $m$ itself. In other words we define the phases by what can be learned about the reference in principle, ignoring the question of explicitly relating $m$ to $\tilde{Q}_m$ for the moment. Nevertheless, we will address this question in Section \ref{sec:coarsegrain} where one measures a macroscopic quantity. 
	
	{ 
		Finally, we caution that being in the QD phase does not mean that Alice can send $\log_2 q$ classical bits of information perfectly to the fraction. In an attempt to do so, Alice may initialize the input qubit in $| j \rangle$ to send the message $j$, $j = 0, \dots, q-1$, instead of coupling the input bit to the reference. The receiver of the message measures the fraction in order to infer the message. It is not hard to see that the  outcome probability $p_m = p_{m,j}$ now depends on the message $j$, as follows: 
		\begin{equation}
			p_{m | j} =  \langle j |  Q_m  |  j \rangle \,,\,j = 0,\dots, q-1\,.
		\end{equation}
		Suppose also that the observer (measuring $F$) has no prior knowledge on the message. Then, by Bayes' theorem, upon obtaining the outcome $m$, the observer may infer that the message is $j$ with probability 
		\begin{equation}
			p_{j|m} = \frac{ p_{m | j}}{p_{m | 0} + p_{m | 1} \dots + p_{m | q-1}} \,,\, j = 0,\dots, q-1 \,.
		\end{equation}
		Therefore, the observer does not know the message with certainty, even if $Q_m$ is a always proportional to a projector, which is the definition of the QD phase. We would need to further require that $Q_m \propto | j \rangle \langle j | $ for some $j$ (which depends on $m$). As we shall see in the concrete models below, the latter condition is realized only in a fine-tuned limit and compromised by small perturbations, whereas a stable QD phase exists according to our (weaker) definition.
	}
	


\subsection{Mutual information} \label{sec:mutual}
We now discuss another quantity describing the correlation between $F$ and $R$, the (bipartite, symmetric) mutual information, and review its well-known relations to the classical correlation. Recall that the mutual information is defined as
\begin{equation}
I(F, R) = S(\rho_R) + S(\rho_F) - S(\rho_{RF}) 
\end{equation}
where $\rho_X$ is the reduced density matrix on $X$ of the Choi state $\vert \Psi_V \rangle$. It is known in general~\cite{discord-zurek} that the mutual information is greater or equal to the Holevo bound:
\begin{equation} \label{eq:inequality}
I(F, R) \ge \chi(F; R)
\end{equation}
for any choice of measurement involved in the right hand side. This result is interpreted as follows: the mutual information measures the total correlation, which includes a classical and a quantum part~\cite{Henderson_2001}. The former is captured by $\chi(F;R)$, the latter is quantified by the ``quantum discord'' $I(F, R) - \chi(F; R)$.  

A consequence of the inequality \eqref{eq:inequality} on the phases of information defined in Section~\ref{sec:probedef} is that, in the QD phase, 
\begin{equation}\label{eq:IFRQD}
I(F, R) \to  \ln q = H(R) \,.
\end{equation}
In other words, the mutual information tends to the amount of injected information. The validity of eq.~\eqref{eq:IFRQD} for arbitrarily small fractions, sometimes known as the ``QD plateau'', is a well-known signature of the establishment of classical objectivity~\cite{zurek-QD}. 

A trivial example where $I > \chi$ is when $F = E = A$ and $V = \mathbf{1}$. Since $F$ and $R$ are in a maximally entangled pure state, $I(F, R) =  2\ln q$. Meanwhile, any strong measurement on $F$ will completely disentangle $R$, so that $\chi(F,R) = \ln q$, which is the maximal possible value. When $q=2$, the origin of this difference is easy to explain: we can measure a maximally entangled pair of spins in either $x$ or $z$ directions and find perfect correlation (this corresponds to $I = 2\ln 2$), but we cannot perform \textit{both} measures at the same time (so $\chi$ can only be $\ln 2$). Thus, such ``quantum'' correlation can be only revealed by experiments of Bell type~\cite{bell1964einstein,EPR-review}, and arguably not relevant for classical objectivity~\cite{girolami,girolami1,Korbicz-rev}. This justifies the choice of defining the phases of information using $\chi$ instead of $I$ in general. 

It is also known~\cite{discord-zurek} that quantum discord vanishes, that is, the equality in \eqref{eq:inequality} holds, if $\pi_m$ are a complete set of one-dimensional projectors and $\rho_{FR}$ is block diagonal in the measurement basis:
\begin{equation} \label{eq:nodiscordcondition}
\rho_{FR} =  \sum_m \pi_m \rho_{FR} \pi_m  = \sum_m  p_m  (| m \rangle \langle m|)_{F} \otimes \rho_m 
\end{equation}
where $| m \rangle$ is a normalized state such that $\pi_m | m \rangle = |m \rangle$. Indeed, if this is the case, we can compute by block
\begin{align*} 
S(FR) =& - \sum_m \mathrm{Tr}[p_m \rho_m \ln (p_m \rho_m) ] \\ 
=&  - \sum_m p_m \mathrm{Tr} [\rho_m \ln (\rho_m)] - 
\sum_m p_m \ln p_m  \\
=&  \sum_m p_m S(\rho_m) + S(F) \,.
\end{align*}
Comparing this \eqref{eq:JFRdef} we see that $\chi(F;R) = I(F;R)$. 

The condition~\eqref{eq:nodiscordcondition} can be realized essentially by applying a dephasing channel to $F$, which removes the block off-diagonal density matrix elements. More specifically, we may do the following: let $V_0$ be any isometry from $\mathcal{H}_A$ to $\mathcal{H}_{E}$. Choose a basis $\{ |m \rangle \}$ for $\mathcal{H}_{F}$ and let 
\begin{equation}
Y_F = \sum_m |m \rangle |m \rangle \langle m |
\end{equation}
be the ``copying'' isometry $\mathcal{H}_{F} \to \mathcal{H}_{F} \otimes  \mathcal{H}_{F'}$ where $F'$ is an identical copy of $F$ ($Y$ acts trivially on $E \setminus F$). We claim that quantum discord vanishes (if we measure the $m$-basis in $\mathcal{F}$): for the CJ state of the amended isometry
\begin{equation} \label{eq:YV0nodiscord}
V = Y_F V_0 \implies   I(F, R) = \chi(F, R)  \,.
\end{equation}
To see why, let
$$\rho_0 = \mathrm{Tr}_{E\setminus F} [\vert \Psi_{V_0} \rangle \langle \Psi_{V_0} |] = \sum_{m,n} (|m \rangle \langle n |)_{F} \, \rho_{R,mn} $$
be the reduced density matrix on ${FR}$ of the CJ state of $V_0$. Then, we see that 
\begin{align*}
\rho_{FR} &= \mathrm{Tr}_{F'} [Y_F \rho_0 Y_F^\dagger]  \nonumber \\ 
& = \sum_{m,n} \mathrm{Tr}_{F'} [ |m \rangle_{F}  |m \rangle_{F'} \langle n  |_{F}  \langle n  |_{F'}] \rho_{R,mn} \nonumber \\
& = \sum_m  |m \rangle_{F} \langle m  |_{F}  \rho_{R,mm} 
\end{align*}
is indeed block diagonal in the $|m \rangle$ basis, as required by \eqref{eq:nodiscordcondition}. We shall see that the structure \eqref{eq:YV0nodiscord} is realized naturally in tree models [see the discussion around~\eqref{eq:nodiscord-micro}]. 

\subsection{The Clifford case}\label{sec:Clifford-gen}
To contrast our general approach with the Clifford-specific method in Ref.~\cite{FC}, let us consider the distribution \eqref{eq:PQdef} when $V$ is deterministic and Clifford: more precisely, this means that there is some Clifford unitary $U$ such that 
\begin{equation}\label{eq:CliffordVdef}
V | \varphi \rangle = U | \varphi \rangle \otimes |  +_z \rangle^{\otimes (|E| - 1)} 
\end{equation}
for all $|\varphi \rangle$ (this formula is the same as \eqref{eq:VfromU} above). We also choose to measure all the qubits in $F$ in the computational basis. We claim that the density matrix distribution $\mathbb{P}(\tilde{Q})$ is either perfectly encoding or is supported on $\{ \mathbf{1} + \sigma^a,  \mathbf{1} - \sigma^a\}$ for some unique $a \in \{x,y,z\}$. 

To see why, let $\mathcal{G} = \left<  \sigma^z_i, i \in F \right>$ be the abelian group generated by the Pauli-$z$'s on $F$. For any $g \in \mathcal{G}$, $U^\dagger g U$ is a product of Pauli's acting on $A$ and the recruits (by the Clifford-ness of $U$), and $V^\dagger g V$ is either $0, \mathbf{1}$ or proportional to a Pauli on $A$. In fact, it is not hard to show that the following is a group homomorphism~\cite{FC}: 
\begin{equation}\label{eq:homo}
\mathcal{G}_0 = \{g\in \mathcal{G}:V^\dagger g V \ne 0\} \ni g \mapsto V^\dagger g V  \in \mathcal{P}_A
\end{equation} 
where $\mathcal{P}_A $ is the Pauli group of $A$. Meanwhile, the set of measurement outcomes can be identified with the dual group of homomorphisms (characters) $m: \mathcal{G} \to \{1, -1\}$, such that 
$$ \pi_m = \frac{1}{|\mathcal{G}|} \sum_{g \in \mathcal{G}} g m(g) \implies Q_m = \frac{1}{|\mathcal{G}|}  \sum_{g \in \mathcal{G}_0} V^\dagger g V m(g). $$
Now, one of the two possibilities must happen:
\begin{itemize}
\item 
If $ V^\dagger \mathcal{G}_0 V \subset \{\pm \mathbf{1}\}$, then for any $m$, $Q_m$ either vanishes or is proportional to $\mathbf{1}$. So $\mathbb{P}(\tilde{Q})$ must be perfectly encoding. 
\item  Otherwise, $V^\dagger \mathcal{G}_0 V $ contains some nontrivial Pauli, $\pm \sigma^a$, $a \in \{ x,y,z \}$. Since $V^\dagger \mathcal{G}_0 V$ is abelian, $a$ is unique. Now, for any $m$, $g \mapsto V^\dagger g V m(g) $ is also a group homomorphism. If its image contains $\{\mathbf{1}, -\mathbf{1}\}$, then $Q_m = 0$ and the measurement outcome $m$ cannot occur. Otherwise, $\{ V^\dagger g V m(g) : g\in \mathcal{G}_0 \} = \{ \mathbf{1}, u \sigma^a \} $ for some $u \in \{1, -1\}$ and we have $$ Q_m = \frac{|G_0|}{|\mathcal{G}|} \frac12 (\mathbf{1} + u \sigma^a) \,,\, $$
which is proportional to a pure state in the $\pm a$ direction. 
\end{itemize}
We thus conclude that in a deterministic Clifford model, $\mathbb{P}(\tilde{Q})$ can only be perfectly encoding or perfectly QD; in the latter case, it is supported on a set of two points, $\{ \mathbf{1} + \sigma^a,  \mathbf{1} - \sigma^a\}$ for some unique $a = x, y$ or $z$. Any intermediate behavior can only be a stochastic mixture resulting from averaging over a random ensemble of models. By doing that, we always obtain a simple distribution $\mathbb{P}(\tilde{Q})$ whose support is a finite set contained in $\{\mathbf{1}\} \cup \{\mathbf{1} \pm \sigma^a: a = x, y, z \}$. This is why Clifford models are simple to solve, see also Section~\ref{sec:clifford-tree} for an example. As we shall see, beyond Clifford, a deterministic model can have qualitatively different intermediate QD phases, where  $\mathbb{P}(\tilde{Q})$ has a continuum support. 

Finally, we note that the analysis here is less general than the Clifford-specific approach of Ref.~\cite{FC}, which allows to probe \textit{both} quantum and classical correlations. To do this in generic, non-Clifford models, we may resort to the mutual information, for which the standard calculation technique requires taking a replica limit. Alternatively, we need to consider Bell-type experiments with measurements that cannot be simultaneously performed. Investigation along those lines will be left to the future.


\section{Expanding tree models}\label{sec:trees}
In this Section we apply the above general formalism to the study of two similar hierarchical models of structured environment. The models will be defined in Section~\ref{sec:treedefs}. In Section~\ref{sec:recursion}, we derive the exact recursion relations satisfied by the density matrix distribution. These apply to the ``microscopic measurement''. It is the strong measurement in the computational basis, for which we shall show that the quantum discord vanishes exactly. So this will be the ``default'' measurement with which we define the phase diagram. Section~\ref{sec:phasediagram} is the technical core of this paper, where we establish the phase diagram and study in detail the critical points. In Section~\ref{sec:coarsegrain} we turn to considering ``coarse-grained'' measurements, which can probe the encoding-intermediate transition, but not the QD-intermediate one. Finally, Section~\ref{sec:clifford-tree} discusses a simple random Clifford version of the models studied so far; curiously, this simple model exhibits a direct QD-encoding transition.

\subsection{Models}\label{sec:treedefs}
To define the models, we will proceed in two steps. First, we construct the isometry $V$ (Section~\ref{sec:isometrydef}). Then we specify the environment fractions and the types of measures applied on them (Section~\ref{sec:fractiondef}). 

\subsubsection{Isometry}\label{sec:isometrydef}
The isometry $V$ is constructed as a tensor network with the geometry of a binary tree. The construction is very similar to that of the random Clifford models of Ref.~\cite{FC}, except that $V$ is not Clifford. We will consider two variants: one is deterministic, and the other has a random isometry. They will lead to similar physical properties, yet each variant proves to have its own technical appeal.

Let us first define the deterministic variant. We will use two basic building blocks: a one-site unitary $U: \mathbb{C}^q \to \mathbb{C}^q $ and a branching isometry $Y: \mathbb{C}^q \to \mathbb{C}^q \otimes \mathbb{C}^q$, to be specified below. Then, we can construct a sequence of isometries 
\begin{equation}
V_n: \mathcal{H}_A= \mathbb{C}^q \to \left(\mathbb{C}^q\right)^{\otimes 2^n} = \mathcal{H}_E , n = 1, 2, 3, \dots \,,
\end{equation}  
recursively as follows:
\begin{align}
V_{1} &= Y \,,\,
V_{n+1} =  (V_{n} \otimes V_n) \hat{Y},  n = 0, 1,2 \,, \label{eq:recursion-words} \\ 
\text{where } \hat{Y} &:=  (U \otimes U) Y \,.  \label{eq:Yhat}
\end{align}
It is convenient to adopt a standard tensor network graphical representation, where 
$$ U = \includegraphics[scale=.8,valign=c]{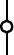}  \;, \; Y = \includegraphics[scale=1,valign=c]{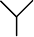} \,, $$ 
and both maps act from bottom to top. Then the recursion relation can be represented as follows:
\begin{equation}\label{eq:recursion-pic}
V_{n+1} = \includegraphics[scale=1,valign=c]{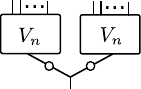} \,.
\end{equation}
Here each isometry maps the bottom qudit to the top qudits. Iterating this, one may readily see that $V_n$ is represented by a tensor network binary tree with $n$ layers of branching vertices, and $2^n$ leaves (output qudits). {   For example, starting from $$ V_1 = Y = \includegraphics[scale=1,valign=c]{Y.pdf} \,, $$
the tensor network of $n = 2$ and $n = 3$ are built as follows:
\begin{equation}\label{eq:V3illus}
	V_2 =   \includegraphics[scale=1,valign=c]{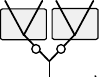} \,,\,
	V_3 =   \includegraphics[scale=1,valign=c]{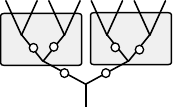} 
\end{equation}
where each box contains a copy of the previous generation, compare to \eqref{eq:recursion-pic}. It is worth noting that the recursion works by adding a layer at the bottom of the tree, which corresponds to early time in terms of the dynamics. This is known as a ``backward recusion'', as opposed to the forward one which adds a layer at the top.} 

It remains to specify $U$ and $Y$ for our models. They are qubit models, so $q=2$, and we will view qubits as spin halves in the standard way. The branching isometry is defined as  
\begin{align}\label{eq:Ydef}
Y = |+_z\rangle |+_z\rangle \langle +_z | + |-\rangle |-_z \rangle \langle -_z|\,,
\end{align}
where $|\pm_z\rangle$ are the eigenstates of $\sigma^z$. The unitary is a rotation:
\begin{equation}\label{eq:Udef}
U = e^{-i \sigma^y \theta / 2} =  \begin{pmatrix}
	\cos \frac\theta2 &  -\sin \frac\theta2 \\ \sin \frac\theta2 &  \cos \frac\theta2 \
\end{pmatrix} 
\end{equation}
where $\sigma^y = \begin{pmatrix}
0 & -i \\ i  & 0
\end{pmatrix}$ is the Pauli-$y$ matrix. We shall parametrize the angle as
\begin{equation}
\theta = J \pi / 2  \quad \text{(deterministic)}
\end{equation}
where $J\in (0,1)$ is the tuning parameter of the model. 

Having constructed the isometry of the deterministic model, we can obtain that of the random model by making the following change: the rotation angle of each one site unitary is now chosen randomly and independently among two opposite values:
\begin{equation}\label{eq:thetarandom}
\theta = \pm J \pi / 2 \quad  \text{(random)} \,,
\end{equation}
with equal probability. We can still define the random isometry $V_n$ recursively: $V_1=Y$ remains deterministic; to generate a realization of $V_{n+1}$, we take two independent realizations of $V_n$, $V_{n,\ell}$ and $V_{n, r}$, and two independent random rotations $U_{\ell}, U_r$ defined by \eqref{eq:Udef} and \eqref{eq:thetarandom}, and let
\begin{equation} \label{eq:Vrecursion-random}
V_{n+1} =  (V_{n,\ell} \otimes V_{n, r}) \, (U_{\ell} \otimes U_r) \, Y  \,, n \ge 1  \,.
\end{equation}

The basic intuition motivating the above definitions is very simple. The branching isometry $Y$ broadcasts the $z$-component of its input to the two outputs (descendants). As a result, when $J = 0$, the CJ state $\Psi_V$ is a GHZ state on $E \cup R$,
\begin{equation}\label{eq:PhiJ0}
|\Phi_V \rangle_{J =0} = \frac1{\sqrt{2}} ( |+_z\rangle_R | +_z \rangle^{\otimes |E|}  + | -_z\rangle_R | -_z \rangle^{\otimes |E|} )  \,,
\end{equation}
and has an ``ideal'' QD behavior: the $z$-component of the injected spin is broadcast to all the environment ones. Now, $Y$ cannot broadcast the $x$-component as well, due to no-cloning. So, when $J$ increases, the one-site rotation perturbs the broadcasting process more and more strongly, with the maximal effect expected at $J = 1$, where the rotation transforms $z$ to $x$. Hence, we may expect a QD phase at small $J$ and an encoding phase at large $J$ (this will be confirmed below). We may also view our models as perturbations of a log-depth circuit generating the repetition code. It is curious to note that log-depth circuits are known as well for the toric code~\cite{vidal-toric}. 

{ 
Before moving on, we recall that the branching isometry $Y$ can be obtained by applying a \texttt{CNOT} gate to the input qubit and a new ``recruit'' qubit in the $ | +_z \rangle = | 0 \rangle $ state~\cite{FC}:
\begin{equation}
	\includegraphics[scale=1.4,valign=c]{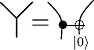} \,.
\end{equation}
Such a \texttt{CNOT} operation, as well as its variants, is a routinely used to model a quantum measurement process, and much discussed in the context of decoherence and emergence of classicality~\cite{zurek-decoherence,girolami,zurek-review,Ryan-onion}. Note also that the recruit qubits are nothing but the initial environment degrees of freedom, see \eqref{eq:VfromU} above. With this in mind, we may represent the $n = 3$ isometry as a circuit:
\begin{equation}
	V_{3} = \includegraphics[scale=1.2,valign=c]{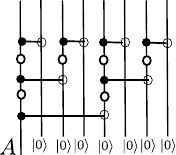}
\end{equation}
We observe that the environment qubits are incorporated gradually into the dynamics, and every pair of qubits interact at most once. In this sense, our models are a type of ``collision model''~\cite{collision-model}, which has already been applied to study quantum Darwinism~\cite{campbell}. In this respect our contribution is a systematic study of phases of information in such models. 
}

\subsubsection{Fraction and measurements}\label{sec:fractiondef}
We now specify the choice of the environment fraction $F \subset E$, where the environment $E$ has $2^n$ bits, represented as the leaves of the binary tree. Note that the information retrievable in $F$ depends not only on its size, but also on how its qubits are distributed with respect to the tree structure. It turns out that to obtain a nontrivial phase diagram, it is necessary to distribute $F$ uniformly among the sub-trees. One way to do this~\cite{FC} is to choose $F$ randomly. Here, we shall consider deterministic fractions $F = F_{t,k}$ with size 
\begin{equation}
|F_{t,k}| = 2^{t}  \,,\, t = n - k \,,\, k \in \{ 2, 3, \dots \} \,,
\end{equation}
such that every sub-tree of size $2^k$ contains exactly one qubit in $F$ (that one qubit can be arbitrarily chosen in the sub-tree, all choices being equivalent). We will often describe $F$ in terms of its relative size, 
\begin{equation}
|F_{t,k}| / |E| = 2^{-k} \le 1/4. 
\end{equation} 
We excluded the case $k = 1, f = 1/2$ for now, since it leads to a qualitatively distinct behavior from $k > 1$, as we shall see below (Section~\ref{sec:IC}). Here is an illustration with $t = 1, k =2$ and $n = t + k = 3$:
\begin{equation} \label{eq:Fexample}
\includegraphics[valign=c,scale=1]{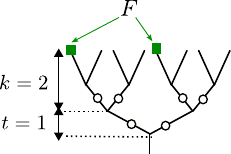} 
\end{equation}
It is hopefully clear from this figure that $F_{t,k}$ can be also constructed by an recursion on $t$. Indeed, $F_{t=0, k}$ contains exactly one qubit, which can be chosen arbitrarily. For $t = 0, 1, 2, \dots$, the fraction $ F_{t + 1,k} $ is obtained by joining two copies of $F_{t,k}$, associated with the two descendants of the root:
\begin{equation}
F_{t + 1,k} = F_{t, k}^{(\ell)} \sqcup  F_{t, k}^{(r)}  \,.
\end{equation}  
In the example \eqref{eq:Fexample} above, the recursive construction of the fraction is illustrated as follows:
\begin{equation}
\includegraphics[valign=c,scale=1]{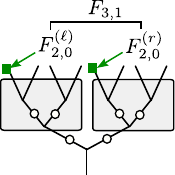} 
\end{equation}

We will consider two type of measurements. First, the basic choice is to measure all the spins in $F$ in the $z$ (computational) basis. This ``microscopic'' measurement will be our focus until Section~\ref{sec:coarsegrain}. In the latter we will consider  ``coarse-grained'' measurements, such as the total magnetization in $F$. 

Let us point out that with the microscopic measurement, the quantum discord vanishes for any $k \ge 1$:
\begin{equation}\label{eq:nodiscord-micro}
I(F,R) = \chi(F,R)  \,,\, k \ge 1 \; \text{(microscopic)} \,.
\end{equation}
Indeed, one can readily see that the isometry $V$ has the form $Y_F V_0$~\eqref{eq:YV0nodiscord} where $Y_F = \otimes_{i\in F} Y_i$ and $V_0$ is the isometry from the root to the second last layer of qubits (adjacent to the output leaves).  

\subsection{Recursion relation}\label{sec:recursion}
The recursive construction of the tree models, explained in the previous section, indicates that the matrix distribution $\mathbb{P}(\tilde{Q})$ can be also computed by a recursion relation. The goal of this section is to derive them for the microscopic measurement. The coarse-grained measurements have slightly different recursion relations, see Section~\ref{sec:coarsegrain} below.

The main idea is to relate the matrix distribution $\mathbb{P}_{t+1,k}(\tilde{Q})$, which involves $V = V_{n+1}$ ($n = t + k$) and $F = F_{t+1,k}$, to $\mathbb{P}_{t,k}(\tilde{Q})$, which involves $V = V_{n}$ and $F = F = F_{t,k}$. This recursion relation allows to increase $t$ and $n$ by $1$ while keeping $k$ fixed. Below, we will first derive general results valid for binary-tree models (Section~\ref{sec:recursion-gen}), where the main result is \eqref{eq:recursion-gen-f}; then we specify to our concrete cases (Section~\ref{sec:recursion-uv}), where the main results are \eqref{eq:recursion-deterministic} and \eqref{eq:recursion-random}. The recursion relations are complemented by the initial condition, $\mathbb{P}_{0,k}(\tilde{Q}) $,  which we will determine in Section~\ref{sec:IC}. We will mainly focus on the deterministic model, where the derivation is more transparent, and amend the procedure to incorporate the randomness. 

\subsubsection{Recursion in general}\label{sec:recursion-gen}
Consider measuring all the qubits in the fraction $F = F_{t+1,k}$ in the $z$-basis. The measurement outcome can be written as $\vec{m} = (m_i)_{i \in F}$ where $m_i = \pm 1$ is the outcome of qubit $i \in F$. The corresponding operators are projectors: 
\begin{equation}
\pi_{\vec{m}} = \prod_{i \in F} \frac{I + m_i \sigma^z_i}2
\end{equation}
where $\sigma^z_i$ is the Pauli-$z$ operator acting on the site $i$. Now observe that we may split the measurement outcomes into two halves coming from the two subtrees of the root: 
\begin{equation}\label{eq:mlmr}
\vec{m} = (\vec{m}_\ell, \vec{m}_r) \,,\, \vec{m}_{\ell,r} = (m_i)_{i \in F_{t,k}^{\ell,r} } \,.
\end{equation} 
Accordingly, the projectors act on the two subtrees in a factorized way:
\begin{equation}
\pi_{\vec{m}} = \pi_{\vec{m}_\ell} \otimes \pi_{\vec{m}_r}   \,.
\end{equation} 
Let us also introduce the time evolution super-operator 
\begin{equation}
\mathcal{L}_{A} (X) := A^\dagger X A \,,
\end{equation}
where $A$ is an isometry. It is routine to check that 
\begin{align}
\mathcal{L}_{U_1 U_2}(X)& = \mathcal{L}_{U_2} (\mathcal{L}_{U_1}(X)) \\
\mathcal{L}_{U_1 \otimes U_2}(X_1 \otimes X_2) &= \mathcal{L}_{U_1}(X_1) \otimes \mathcal{L}_{U_1}(X_2) 
\end{align}   
Then, using the definition~\eqref{eq:Qmdef} and the recursion relation for $V_n$ \eqref{eq:recursion-words}, we have, 
\begin{align}
Q_{\vec{m}} &=  \mathcal{L}_{V_{n+1}} (\pi_{\vec{m}})   \nonumber  \\
&  = \mathcal{L}_{ \hat{Y}} (\mathcal{L}_{V_{n} \otimes V_n} (\pi_{\vec{m}_\ell} \otimes   \pi_{\vec{m}_r}) ) \nonumber \\ 
&= \mathcal{L}_{\hat{Y}} \left( \mathcal{L}_{V_{n}} (\pi_{\vec{m}_\ell}) \otimes  \mathcal{L}_{V_{n}}(\pi_{\vec{m}_r}) \right)  \nonumber \\
& =\mathcal{L}_{ \hat{Y} }  (Q_{\vec{m}_\ell} \otimes Q_{\vec{m}_r} ) \,. \label{eq:Qrecursion}  
\end{align}
Now, recall from \eqref{eq:pmdef} that $ Q_{\vec{m}} = p_{\vec{m}} \tilde{Q}_{\vec{m}}$ where $ \mathrm{Tr}[\tilde{Q}_{\vec{m}}] = q$, and similarly for $\vec{m}_\ell$ and $\vec{m}_r$. Then the above formula can be written as  
\begin{equation}
p_{\vec{m}} \tilde{Q}_{\vec{m}} =  {p_{\vec{m}_\ell} p_{\vec{m}_r}}{} \mathcal{L}_{ \hat{Y}}  (\tilde{Q}_{\vec{m}_\ell} \otimes \tilde{Q}_{\vec{m}_r} )  \,.
\end{equation}
Taking the trace on both sides we find 
\begin{align}
p_{\vec{m}}  &=  {p_{\vec{m}_\ell} p_{\vec{m}_r}} \varphi(\tilde{Q}_{\vec{m}_\ell}, \tilde{Q}_{\vec{m}_r} ), \label{eq:pmrecursion} \\ \text{where }& \varphi(A,B) := \frac1q \mathrm{Tr}\left[ \mathcal{L}_{ \hat{Y}} (A\otimes B) \right] \,, \label{eq:varphidef}
\end{align}
and also 
\begin{align}
\tilde{Q}_{\vec{m}} &=  \mu(\tilde{Q}_{\vec{m}_\ell}, \tilde{Q}_{\vec{m}_r}), \label{eq:Qtilderecursion}  \\
\text{where } \mu(A, B)& :=  \frac{\mathcal{L}_{ \hat{Y} }  (A \otimes B )}{\mathrm{Tr}[\mathcal{L}_{ \hat{Y}}  A \otimes B )] / q} \,. \label{eq:mudef}
\end{align}
The formulas~\eqref{eq:pmrecursion}-\eqref{eq:mudef} are important. They show that the measurement results in the two sub-trees are {correlated}, and that $  \tilde{Q}_{\vec{m}}$ is a nonlinear function of $\tilde{Q}_{\vec{m}_\ell}$ and $ \tilde{Q}_{\vec{m}_r} $. These features make the model nontrivial. 

We are now ready to derive the recursion relation for the density matrix distribution $\mathbb{P}_{t,k}(\tilde{Q})$ in a general form (below we will omit $k$ for brevity). It is more convenient to work with an ensemble average with an arbitrary test function $f(\tilde{Q})$. Recalling the definition \eqref{eq:observ}, and using \eqref{eq:mlmr}, \eqref{eq:pmrecursion} and \eqref{eq:Qtilderecursion}, we have 
\begin{align} \label{eq:recursion-gen-f}
& \left< f(\tilde{Q}) \right>_{\mathbb{P}_{t+1}}   =  \sum_{\vec{m}} p_{\vec{m}} f(\tilde{Q}_m) \nonumber \\ 
= & \sum_{\vec{m}_\ell} \sum_{\vec{m}_r}  {p_{\vec{m}_\ell} p_{\vec{m}_r}} \varphi(\tilde{Q}_{\vec{m}_\ell}, \tilde{Q}_{\vec{m}_r} ) f( \mu(\tilde{Q}_{\vec{m}_\ell}, \tilde{Q}_{\vec{m}_r} ))  \nonumber \\ 
= & \left<  \varphi(\tilde{Q}_{\ell}, \tilde{Q}_{r}) f( \mu(\tilde{Q}_{\ell}, \tilde{Q}_{r} )) \right>_{\mathbb{P}_{t}^{\otimes 2}}
\end{align}
In the average of the last line, $\tilde{Q}_{\ell}$ and $\tilde{Q}_{r} $ are independent and identically distributed as $\mathbb{P}_{t}$. The above formula, valid for any $f$, determines completely the distribution $ \mathbb{P}_{t+1}$, and thus the recursion relation. An explicit (but less useful) formula for $ \mathbb{P}_{t+1}$ is the following:
\begin{align}\label{eq:recursion-gen-P}
&\mathbb{P}_{t+1}(\tilde{Q}) = \nonumber \\ & \int_{\tilde{Q}_\ell , \tilde{Q}_r} \delta(\tilde{Q} - \mu(\tilde{Q}_{\ell}, \tilde{Q}_{r} ))  \varphi(\tilde{Q}_{\ell}, \tilde{Q}_{r} )  \mathbb{P}_{t}(\tilde{Q}_\ell) \mathbb{P}_{t}(\tilde{Q}_r) \,.
\end{align}
The above recursion relation holds for any binary-tree tensor network models where each branching corresponds to a deterministic isometry $\hat{Y}$, and thus, to our deterministic model. Our random model falls into a category where the $\hat{Y}$'s on each vertex are independent, see \eqref{eq:Vrecursion-random}. For these, using the adapted definition of $\mathbb{P}(\tilde{Q})$ \eqref{eq:PRrandom} which also averages over randomness, it is not hard to see that \eqref{eq:recursion-gen-f} still holds provided we also average over $\hat{Y}$ (which affects $\mu$ and $\varphi$) in the right hand side: 
\begin{equation} \label{eq:recursion-gen-f-random}
\left< f(\tilde{Q}) \right>_{\mathbb{P}_{t+1}}  = 
\mathbb{E}_{\hat{Y}}  \left<  \varphi(\tilde{Q}_{\ell}, \tilde{Q}_{r}) f( \mu(\tilde{Q}_{\ell}, \tilde{Q}_{r} )) \right>_{\mathbb{P}_{t}^{\otimes 2}} \,.
\end{equation}
Below, for brevity, we will absorb $ \mathbb{E}_{\hat{Y}} $, into the symbol $\left< [\dots] \right>$:
$$ \left<  [\dots ]\right>_{\mathbb{P}_{t}^{\otimes 2}} :=  \mathbb{E}_{\hat{Y}} \left<  [\dots ]\right>_{\mathbb{P}_{t}^{\otimes 2}} \,.$$

Eq.~\eqref{eq:recursion-gen-f} and \eqref{eq:recursion-gen-f-random} are the main result of this section, and valid for general binary tree models. (Generalization to general trees is straightforward to write down, but we will not need that.) Before applying them to our specific qubit models, let us discuss a few simple general properties. 

First let us check that the recursion relation preserves the normalization $\left< 1 \right> = 1$ and the property  $\left< \tilde{Q} \right> = \mathbf{1}$~\eqref{eq:Qaverageis1} that we know to hold for any density matrix distribution. Let us proceed by induction, assuming that $\left<1 \right>_{\mathbb{P}_{t}} = 1$ and $\left< \tilde{Q}_{\ell, r} \right>_{\mathbb{P}_{t}} = \mathbf{1}$. Then, setting $f=1$ in \eqref{eq:recursion-gen-f}, we have 
\begin{align}
&\left< 1 \right>_{\mathbb{P}_{t+1}} = \left< \frac1q \mathrm{Tr}[\mathcal{L}_{\hat{Y}}(\tilde{Q}_{\ell} \otimes \tilde{Q}_{r})] \right>_{\mathbb{P}_{t}^{\otimes 2}} \nonumber \\  
& =  
\frac1q \mathrm{Tr}\left[\mathcal{L}_{\hat{Y}}\left(\left< \tilde{Q}_{\ell} \right>_{\mathbb{P}_{t}} \otimes \left<\tilde{Q}_{r}\right>_{\mathbb{P}_{t}}\right)\right] \nonumber \\
& =  \frac1q \mathrm{Tr}\left[\mathcal{L}_{\hat{Y}} \left( \mathbf{1} \otimes \mathbf{1} \right)\right] = \frac1q \mathrm{Tr}\left[\mathbf{1}\right]  = 1 \,,
\end{align}
where the last line follows from the isometry of $\hat{Y}$.  This equation shows that the recursion relation provides a correctly normalized distribution. Similarly, we have 
\begin{align}
\left< \tilde{Q} \right>_{\mathbb{P}_{t+1}} = \left< \mathcal{L}_{\hat{Y}}(\tilde{Q}_{\ell} \otimes \tilde{Q}_{r}) \right> = \mathbf{1} \,,
\end{align}
which means that the property $\left< \tilde{Q}_{\ell, r} \right> = 1$ is preserved by the recursion. 

Next, we note that the perfectly encoding distribution $\mathbb{P}(\tilde{Q}) = \delta(\Tilde{Q} - \mathbf{1})$ is a fixed point of the recursion map. 
Indeed, the isometry of $\hat{Y}$ implies $\mathcal{L}_{\hat{Y}}(\mathbf{1}\otimes \mathbf{1}) = \mathbf{1}$ so $\mu(\mathbf{1}\otimes \mathbf{1}) = \mathbf{1} $. Thus if $\tilde{Q}_{\ell,r} = \mathbf{1}$ almost surely, so does $\tilde{Q}$. Finally, a ``perfectly QD'' distribution in which $\tilde{Q} / q$ is almost always a projector is sent to another perfectly QD distribution by the recursion map. This is because if $\tilde{Q}_{\ell,r}\propto | \psi_{\ell,r} \rangle\langle \psi_{\ell,r} |$, then 
$$\tilde{Q} \propto \mathcal{L}_{\hat{Y}}(\tilde{Q}_{\ell} \otimes \tilde{Q}_{r}) \propto     |\psi \rangle \langle \psi| \,,\, |\psi \rangle = Y^\dagger (| \psi_{\ell} \rangle \otimes |\psi_{\ell} \rangle) $$
which is also proportional to a projector. For later reference, let us summarize the invariance of perfectly QD and encoding distributions in terms of the purity~\eqref{eq:puritydef}:
\begin{align}\label{eq:r2fixedpoints}
\forall b \in \{0, 1\} ,  \left< r^2 \right>_{t} = b \implies   \left< r^2 \right>_{t+1} = b \,.
\end{align}

\subsubsection{Recursion in concrete models}\label{sec:recursion-uv}
We now apply the above general formulas to our concrete models. Since the building blocks of our models, $U$~\eqref{eq:Udef} and $Y$~\eqref{eq:Ydef}, are represented by real matrices in the computational basis, all the $\tilde{Q}$ matrices will be real symmetric with trace $q=2$. So we can parametrize them by a real 2D vector $(u,v)$ as follows:
\begin{equation} \label{eq:defparamuv}
\tilde{Q} = \mathbf{1} + u \, \sigma^z + v \, \sigma^x  \,.
\end{equation}
Similarly, $\tilde{Q}_{\ell,r}$ are parametrized by $(u_{\ell,r}, v_{\ell, r})$. Note that the purity~\eqref{eq:puritydef} equals the squared norm of the vector:
\begin{equation}
r^2 = \frac{ \mathrm{Tr}[\tilde{Q}^2] - q}{q^2 - q} = u^2 + v^2 \,.
\end{equation} 
Also, as $\tilde{Q}/q$ is a density matrix, $u^2 + v^2 = r^2 \le 1$. So, any density matrix $\mathbb{P}(u,v)$ is supported in the unit disk. A perfectly QD distribution is supported on the unit circle $u^2 + v^2 = 1$, and the perfectly encoding one is peak at the origin $u = v = 0$.  Finally, the property $\left<  \tilde{Q}\right> = \mathbf{1}$~\eqref{eq:Qaverageis1} translates to
\begin{equation}
\left< u \right> = \left< v \right> = 0 \label{eq:uvmeanis0}
\end{equation}
in any ensemble. 

Let us now compute $ \mathcal{L}_{\hat{Y}} (\tilde{Q}_{\ell} \otimes \tilde{Q}_{r})$. Since $\hat{Y}  = (U_\ell \otimes U_r)\otimes Y$~\eqref{eq:Yhat}, we have 
\begin{equation}
\mathcal{L}_{\hat{Y}} (\tilde{Q}_{\ell} \otimes \tilde{Q}_{r}) = 
\mathcal{L}_Y \left( \mathcal{L}_{U_\ell}(\tilde{Q}_{\ell} )\otimes \mathcal{L}_{U_r}(\tilde{Q}_{r}) \right)  \,.
\end{equation}
As $U_{\ell,r} = e^{-i \theta_{\ell,r} \sigma^y / 2}$~\eqref{eq:Udef} is a rotation, we have (recall $[\sigma^y, \sigma^z] = 2 i \sigma^x$ and $[\sigma^y, \sigma^x] = - 2 i \sigma^z $):
\begin{equation}
\tilde{Q}_{\ell,r}' := \mathcal{L}_{U_{\ell,r}} (\tilde{Q}_{\ell,r}) = \mathbf{1} + u'_{\ell,r} \sigma^z +  v'_{\ell,r} \sigma^z  
\end{equation}
where
\begin{equation} \label{eq:rotation}
\begin{pmatrix}
	u'_{\ell,r} \\ v'_{\ell,r}
\end{pmatrix} = \begin{pmatrix}
	\cos\theta_{\ell,r}  & - \sin\theta_{\ell,r} \\  \sin\theta_{\ell,r}  &  \cos\theta_{\ell,r}
\end{pmatrix} \begin{pmatrix}
	u_{\ell,r} \\ v_{\ell,r}
\end{pmatrix} 
\end{equation}
are the vectors rotated by $\theta_{\ell,r}$ (we will routinely use the prime to denote the rotation in order to save space). Now, since $Y = \sum_{i} | i i \rangle \langle i | $~\eqref{eq:Ydef} in the $z$ basis, the action of $\mathcal{L}_Y $  amounts to an ``element-wise'' multiplication of the matrix elements in the $z$-basis (computational basis):
\begin{equation}
\langle i | \mathcal{L}_Y (A \otimes B) |  j \rangle=  \langle i| A |j \rangle \langle i| B |j \rangle \,.
\end{equation}
In this basis $\mathbf{1} + u \sigma^z + v \sigma^x = \begin{pmatrix}
1 + u & v \\ v& 1-u
\end{pmatrix}$, so we have 
\begin{align}
&\mathcal{L}_{{Y}} (\tilde{Q}'_{\ell} \otimes \tilde{Q}'_{r}) = \begin{pmatrix}
	(1 + u'_\ell)  (1 + u'_r)  & v'_\ell v'_r \\  v'_\ell v'_r & (1 - u'_\ell)  (1 - u'_r)
\end{pmatrix} \nonumber  \\ 
=& (1 +  u'_\ell  u'_r) \left( \mathbf{1} +  \frac{u'_\ell+ u'_r}{1 + u'_\ell u'_r} \sigma^z +  \frac{v'_\ell v'_r}{1 + u'_\ell u'_r}  \sigma^x \right) \,.
\end{align}
This gives following explicit formulas for the functions $\mu$~\eqref{eq:mudef} and $\varphi$~\eqref{eq:varphidef}:
\begin{align}
& \varphi( (u_\ell, v_\ell ),   (u_r, v_r ) ) = 1 + u'_\ell u'_r  \,, \label{eq:varphi-model} \\ 
& \mu((u_\ell, v_\ell ),   (u_r, v_r )) = \left( \frac{u'_\ell+ u'_r}{1 + u'_\ell u'_r},   \frac{v'_\ell v'_r}{1 + u'_\ell u'_r} \right) \,. \label{eq:mu-model}
\end{align}
Now, in the deterministic model, the rotation angle is deterministic $\theta_{\ell,r} = J\pi / 2$, so \eqref{eq:varphi-model}-\eqref{eq:mu-model} lead immediately to the following recursion relation:
\begin{align}
&\left< f(u,v) \right>_{\mathbb{P}_{t+1}} = \label{eq:recursion-deterministic} \\  & \left<  (1 + u'_\ell u'_r) 
f\left(\frac{u'_\ell+ u'_r}{1 + u'_\ell u'_r}, \frac{v'_\ell v'_r}{1 + u'_\ell u'_r} \right)   \right>_{\mathbb{P}_{t}^{\otimes 2}, \theta_{\ell,r} = J\pi/2}  \nonumber  
\end{align}
Here, $u'_{\ell,r}, v'_{\ell,r}$ are given by \eqref{eq:rotation} with $\theta_{\ell,r}=J\pi/2$. In the right hand side average, $(u_\ell,v_\ell)$ and $(u_r,v_r)$  are independent and identically distributed as $\mathbb{P}_t$.

In the random model, we need to further average over $\theta_{\ell,r} = \pm J \pi / 2$ (the four possibilities have probability $1/4$ each):
\begin{align}
&\left< f(u,v) \right>_{\mathbb{P}_{t+1}} =\label{eq:recursion-random}  \\  &  \left<  (1 + u'_\ell u'_r) 
f\left(\frac{u'_\ell+ u'_r}{1 + u'_\ell u'_r}, \frac{v'_\ell v'_r}{1 + u'_\ell u'_r} \right)   \right>_{\mathbb{P}_{t}^{\otimes 2}, \theta_{\ell,r} = \pm \frac{J \pi}2 } \nonumber \,.
\end{align}
Equations~\eqref{eq:recursion-deterministic} and \eqref{eq:recursion-random} are the main results of this section. They will be analyzed numerically and analytically in Section~\ref{sec:phasediagram}. 

\subsubsection{Initial condition}\label{sec:IC}
The recursion relations \eqref{eq:recursion-deterministic} and \eqref{eq:recursion-random} need to be complemented by the initial condition $\mathbb{P}_{t=0, k}$, corresponding to a  single qubit fraction $|F| = 1$. So the distribution $\mathbb{P}_{t=0, k}$ consists of two delta peaks (at least before averaging over randomness). To find them, it suffices to compute the operator $Q_m^{(k)}$ for the two measurement outcomes $m = \pm 1$. We can do this by a recursion in $k$. Indeed, the recursion construction of the isometry \eqref{eq:recursion-words} implies the following:
\begin{align}
& Q_m^{(1)} = \mathcal{L}_{{Y}} \left( \frac{\mathbf{1} + m \sigma^z }2 \otimes \mathbf{1} \right) = \frac{\mathbf{1} + m \sigma^z }2 \,. \\ 
& Q_m^{(k+1)} = \mathcal{L}_{\hat{Y}} \left( Q_m^{(k)} \otimes \mathbf{1} \right) \,.
\end{align}
Plugging in the explicit definition of $\hat{Y}$, it is not hard to find that:
\begin{equation}\label{eq:Qminit}
Q_m^{(k)} = \frac{\mathbf{1} + m c^{k-1} \sigma^z}2 \,,\, c := \cos(J\pi/2) \,.
\end{equation}
Note that since $\cos(\theta) = \cos(-\theta)$, the above holds in the deterministic model and any instance of the random model, since $\theta = \pm J\pi/2$ in all cases. Therefore, for both models, the initial condition for the recursion relation is 
\begin{equation}\label{eq:P0k}
\mathbb{P}_{0, k}(u,v) = \frac12 \sum_{m = \pm 1} \delta(u - m \, c^{k-1}) \delta(v)  \,,
\end{equation}
in terms of the parametrization \eqref{eq:defparamuv}. We observe that the initial condition \eqref{eq:P0k} depends explicitly on the relative fraction size $|F|/|E|= 2^{-k}$ [by contrast, the recursion relations \eqref{eq:recursion-deterministic} and \eqref{eq:recursion-random} do not depend on $k$]. In particular, $\mathbb{P}_{0, k}$ is perfectly QD when and only when $k = 1$; in that case, $ \mathbb{P}_{t, k}$ will be perfectly QD for $t$, regardless of $J$, see above~\eqref{eq:r2fixedpoints}. This is why $k=1$ is special and shall be considered separately from $k > 1$. In the latter case, $ \mathbb{P}_{0, k}$ is neither perfectly QD nor perfectly encoding; it tends to being perfectly encoding when $k \to \infty$, that is, when the relative fraction size  $|F|/|E| \to 0$. Depending on $J$, it can flow to a QD, encoding or intermediate distribution as $t\to\infty$ under the iterated action of the recursion map. The asymptotic behavior of the recursion flow determines the phase diagram, and will be our main focus in the next Section.

\subsection{Phase diagram (microscopic measurement)}\label{sec:phasediagram}
In this section we study the phase diagram of the models, by analyzing the recursion relations we just derived, with a combination of numerical and analytical techniques. After discussing the numerical methods in Section~\ref{sec:numerics}, we will present an overview of the phase diagram in Section~\ref{sec:overview}, followed by a short discussion on initial-condition independence (Section~\ref{sec:ICindependence}). Then the encoding-intermediate and QD-intermediate transitions will be studied in detail, in Section~\ref{sec:encoding} and \ref{sec:QD} respectively. We will discuss an application to redundancy (Section~\ref{sec:redundancy}) before a brief conclusion.

\begin{figure*}
\centering
\includegraphics[scale=1]{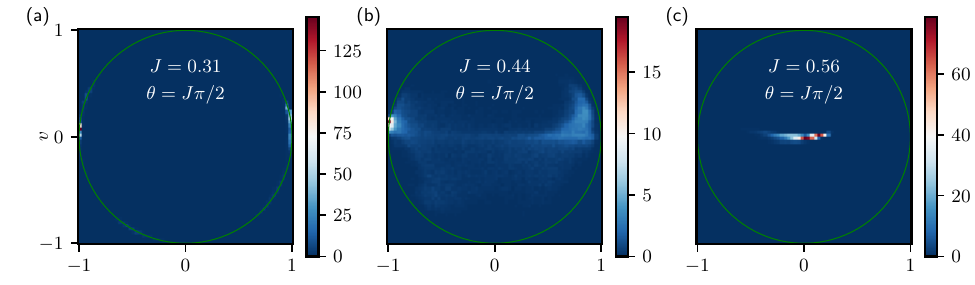}
\includegraphics[scale=1]{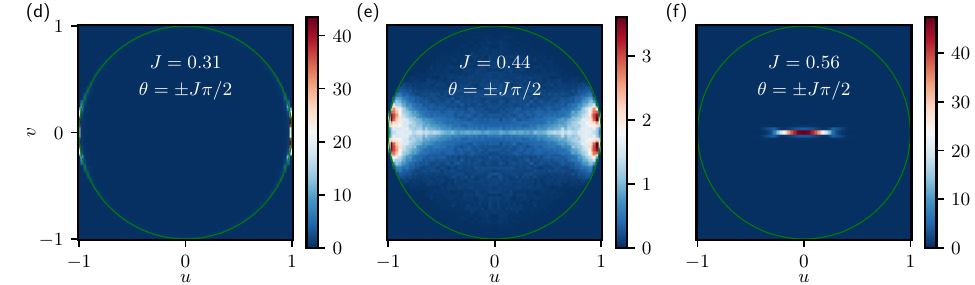}
\caption{The matrix distribution $\mathbb{P}_{t,k}(\tilde{Q}) = \mathbb{P}_{t,k}(u,v) $ where $\tilde{Q} = \mathbf{1} + u \sigma^z + v \sigma^x$, for three values of the parameter $J$ in the deterministic (a-c) and random (d-f) tree models. The distributions are obtained by numerically solving the recursion relations~\eqref{eq:recursion-deterministic} and \eqref{eq:recursion-random}, by methods in Section~\ref{sec:numerics}. These distributions describe the injected information retrievable from a fraction of size $|F| = 2^t$ in an environment of size $|E| = 2^{t + k}$; here $t = 7$ and $k = 2$. (a,d) When $J$ is close enough to $0$, the distribution is supported on the unit circle $u^2+v^2 = 1$ (indicated in green); this indicates a QD phase where the injected information is entirely retrievable from $F$. (b,e) For intermediate values of $J$, the distribution has a nontrivial shape supported throughout the unit disk. The injected information is partially accessible from $F$. (c,f) When $J$ is close enough to $1$, the distribution is peaked at $u = v = 0$. This is the hallmark of an encoding phase where no information is accessible from $F$. }
\label{fig:Ps}
\end{figure*}
\subsubsection{Numerical methods}\label{sec:numerics}
We start by discussing numerical implementations of the recursion relations, which are nontrivial; the impatient reader may skip to Section~\ref{sec:overview} for numerical results. 

First, the brute-force method consists in representing $\mathbb{P}_{t}$ exactly, as a sum of delta peaks:
\begin{equation}
\mathbb{P}_{t} = \sum_i p_i \delta(u_i, v_i) \,,
\end{equation}
where we used the shorthand $\delta(x,y) := \delta(u-x) \delta(v-y)$. To obtain $\mathbb{P}_{t+1}$, we may perform two steps:
\begin{enumerate}
\item \textit{Rotation}: In the deterministic variant, we rotate all $ (u_i, v_i) $ by $\theta = J \pi / 2$. In the random variant, we make two copies of each $(u_i, v_i)$ (with weight $p_i/2$ each), and rotate them by $\pm J \pi / 2$ respectively. Let the intermediate result be $ \sum_i p'_i \delta(u-u'_i) \delta(v-v'_i)$. 
\item \textit{Branching}: We calculate the effect of the branching isometry exactly:
\begin{equation} \label{eq:branching-num}
	\mathbb{P}_{t + 1} =  \sum_{ij} p'_i p'_j ( 1 + u'_i u'_j) \delta\left( \frac{u'_i + u'_j}{1 + u'_i u'_j}, \frac{v'_i v'_j}{1 + u'_i u'_j} \right) \,.
\end{equation}
\end{enumerate}
This method is straightforward and exact. But the number of delta peaks grows double-exponentially in $t$, or exponentially in the fraction size $|F|$: so the brute-force method is no better than an exact representation of the wave-function. In practice we only use it to obtain exact solutions up to $t = 5$ for the deterministic model and $t=4$ for the random one.

To go beyond, we may represent the distribution approximately, as $M$ delta peaks at $(u_i, v_i)_{i=1}^M$ with equal weight $1/M$, where $M$ does not increase with $t$. Such a ``compressed'' distribution can be obtained from an exact one, or another compressed one, by sampling $M$ random points with the appropriate weights. With this in mind, we propose the following approximate algorithm for the deterministic model, where $M = N^2$ is a perfect square:
\begin{enumerate}
\item \textit{Rotation}: The same as the brute-force method above. 
\item \textit{Branching}: Compress the intermediate result to size $N$, calculate the RHS of \eqref{eq:branching-num}, and compress the result (of size $M$) to size $N$. Repeat this $N$ times, and let $\mathbb{P}_{t+1}$ be the sum of the results, which again have $M$ peaks. 
\end{enumerate}
This algorithm is exact in the $N\to\infty$ limit. In practice, $N \in [10^2, 10^3]$ appear to be a good compromise between precision and speed. We can reliably reach $t = 15$ (corresponding to $|F| \ge 10^4$) before the algorithm becomes numerically unstable. 

Finally, there is a faster algorithm, which we shall apply to the random model. To explain its idea, observe that we can rewrite the recursion relation as follows:
\begin{align}
\left< f \right>_{t+1} &= \left< (1 + u'_\ell u'_r) f \right>_t  \nonumber  \\
&= \frac12 \sum_{s = \pm 1} \left< (1 + s u'_\ell) (1 + s u'_r) f \right>_t \,.
\end{align}
This means that, in order to sample an instance $(u,v)$ from the distribution $\mathbb{P}_{t+1}$, we may first draw a random variable $s = \pm 1$ (with equal probability), and then sample $( u'_\ell, v'_\ell) $ and $(u'_r, v'_r)$ \textit{independently} from the randomly rotated ensemble $\{(u', v' )\}$: the probability of picking $(u', v')$ is that given by $\mathbb{P}_t$, multiplied by a bias factor $(1 + s u')$. Moreover, we can show (for example by induction) that in the random model, $\mathbb{P}_t(u,v)$ is has a $\mathbb{Z}_2^2$ under $u \to -u$ and $v \to -v$. Exploiting (and preserving) this symmetry, we end up with the following procedure:
\begin{enumerate}
\item \textit{Rotation}: proceed as in the brute-force method (random case), and obtain the intermediate ensemble $\sum_{i\le 2M} \delta(u'_i, v'_i) / (2M)$.
\item \textit{Branching}: Sample $M$ delta peaks from the biased distribution $\sum_{i\le 2M} (1 + u'_i)  \delta(u'_i, v'_i) / (2M) $, and call them $\{(u'_{i\ell}, v'_{i\ell}\}_{i\le M/2}$. Sample independently $\{(u'_{ir}, v'_{ir}\}_{i\le M/2} $ in the same way. Then let 
\begin{equation} \label{eq:branching-num2}
	\mathbb{P}_{t + 1, \pm} =  \frac{2}{M} \sum_{i \le M} \delta\left( \pm \frac{u'_{i\ell} + u'_{ir}}{1 + u'_{i\ell} u'_{ir}}, \frac{v'_{i\ell} v'_{ir}}{1 + u'_{i\ell} u'_{ir}} \right)  \,.
\end{equation}
Finally let $ \mathbb{P}_{t + 1} = \frac12  \sum_{s = \pm} \mathbb{P}_{t + 1, s}$.
\end{enumerate}
The trick in the branching step is that we only sample explicitly for $s = +1$, and then obtain a (correctly biased) sample for $s = -1$ by symmetry. Also, we force the samples to come in pairs $(u,v), (u, -v)$. As a result, the procedure preserves the $\mathbb{Z}_2^2$ symmetry, and thus the property $\left< u \right> = \left< v \right> = 0$~\eqref{eq:uvmeanis0}. We observed empirically that preserving this property stabilizes the approximate algorithm, and allows us to reach $t \sim 10^2$ in the random model; with $M = 10^4 \sim 10^5$, the algorithm is reasonably fast and precise. (By contrast the above algorithm turns out not stable enough for the deterministic model in order to outperform the slower algorithm above. The culprit is presumably that the matrix distribution of deterministic model does not have the $\mathbb{Z}_2^2$ symmetry; hence, preserving $\left< u \right> = \left< v \right> = 0$ is nontrivial.)

\subsubsection{Phase diagram overview}\label{sec:overview}
Using the above methods, we calculated the density matrix distributions $\mathbb{P}_{t}$ up to $t = 10$ (corresponding to a fraction of size $|F| = 1024$ qubits!) for different values of the parameter $J \in (0,1)$ and initial condition $k>1$; recall that the latter is related to the relative fraction size, $|F|/|E| = 2^{-k}$. 

In Fig.~\ref{fig:Ps} we plot the obtained distribution in the $(u,v)$ plane with three representative values of $J$ for both models. As we anticipated, the distributions of the random model are symmetric with respect to $u \to -u$ and $v\to-v$, while those of the deterministic model do not have any visible  symmetries. Despite this difference, in both models, the distribution becomes more concentrated near $u=v=0$ as $J$ approaches $1$. In contrast, as $J$ approaches $0$, the distribution becomes more concentrated on the unit circle, and more particularly, near the the poles $u = \pm 1, v = 0$. Finally, for intermediate values of $J$, the distributions has a nontrivial shape, covering a finite portion of the of the unit disk. 

\begin{figure}
\centering
\includegraphics[scale=1]{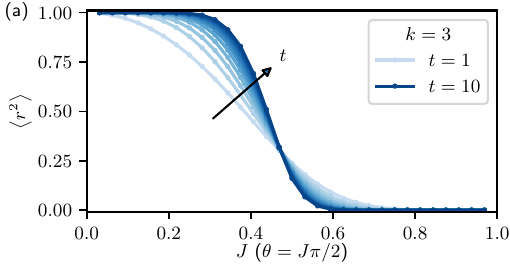}
\includegraphics[scale=1]{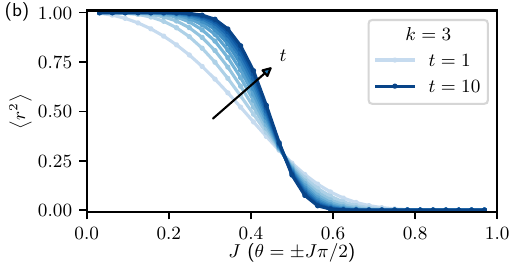}
\caption{The purity $\left< r^2 \right> = \left< u^2 + v^2 \right>$ of the density matrix distribution $\mathbb{P}_{t,k} $as a function of the parameter $J$ for different values of $t$ and $k = 3$ (the fraction size is $|F| = 2^t$ and its relative size is $|F|/|E| = 2^{-k}$), in the deterministic (a) and random (b) tree models. In both models, $\left< r^2 \right> \to 1$ rapidly as $t$ increases when $J$ is small enough, and $ \left< r^2 \right> \to 0$ rapidly when $J$ is close enough to $1$. At intermediate values, the convergence to the thermodynamics limit is slow. Further analysis (see below) will show that an intermediate phase where $0 < \left< r^2 \right>_{t\to\infty} < 1$ separates the QD phase and the encoding phase. The data are obtained by numerically solving the recursion relations~\eqref{eq:recursion-deterministic} and \eqref{eq:recursion-random}, see Section~\ref{sec:numerics} for methods.}
\label{fig:basic}
\end{figure}
To start probing quantitatively the phase diagram, we measure the purity of the distribution $\left<r^2\right>= \left< u^2  + v^2 \right>$. The results are plotted in Fig.~\ref{fig:basic} as a function of $J$, and for different values of $t$ (fraction size). As we may expect, the deterministic and random models yield nearly identical qualitative behaviors. For $J$ close enough to $0$, we observe that $\left< r^2  \right> \to 1$ rapidly as $t \to \infty$, indicating the existence of a stable QD phase at small $J$. Similarly, an encoding phase, characterized by $ \left< r^2  \right> \to 0$, emerges at a region of $J$ close to $1$. The phase diagram at intermediate $J \approx 0.5$ is a priori not clear from Fig.~\ref{fig:basic}, due to strong finite size effects. At this stage, the data are consistent with (at least) two possibilities: (1) a direct discontinuous transition from QD to encoding phase and (2) an intermediate phase, and two continuous transitions. A main goal of what follows is to show that (2) actually takes place in our models, by a combination of analytical and numerical arguments. 

\subsubsection{Fraction size independence}\label{sec:ICindependence}
Before proceeding to the detailed analysis of the phase diagram, let us discuss the dependence on the relative fraction size $|F|/|E| = 2^{-k}$. As we have shown, $k$ only affects the initial condition of the recursion, but not the recursion relation itself. Now, the phase diagram is determined by the thermodynamic ($t \to\infty$) limit of $\mathbb{P}_{t,k}$, or, in other words, the attractive fixed points of the recursion flow. Under the reasonable assumption that such a fixed point is unique (and that there is no exotic asymptotic behaviors such as a limit cycle), we expect the $t\to\infty$ limit of $\mathbb{P}_{t,k} $ to be independent of the initial condition, and the phase diagram to depend only on the parameter $J$, not on the fraction size. 

\begin{figure}
\centering
\includegraphics{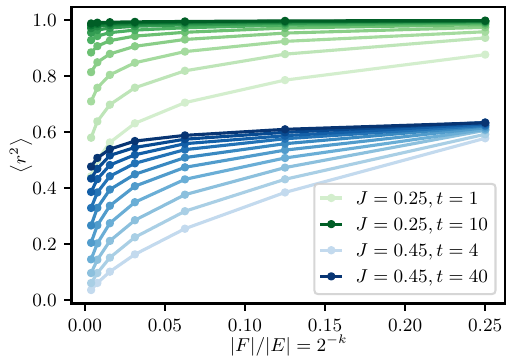}
\caption{Dependence of the purity on the relative fraction size $|F|/|E| = 2^{-k}, k = 2, \dots 8$, for two values of $J$ and a few values $t$ ($|F| = 2^{t}$). At small enough $J$, the purity converges rapidly to a Darwinism plateau $\left< r^2 \right> = 1$ for all values of $k$. At intermediate $J$, the convergence is slower, yet $\left< r^2 \right>$ tends to a $k$-independent value between $0$ and $1$ in the thermodynamic limit.}
\label{fig:kdependence}
\end{figure}
To provide some numerical evidence for the above assumption, in Fig.~\ref{fig:kdependence} we plot the fraction size dependence of the purity for two values of $J$ in the random model. Note that we consider only $k > 1$, since $\left< r^2 \right> =1$ for $k=1$ and any $t$,  see \eqref{sec:IC} above. For the small value $J = 0.25$ (which is deep inside the QD phase, as is apparent in Fig.~\ref{fig:basic} ; see also below), the purity converges to $1$ rapidly for all fraction size. As a consequence the mutual information $I(F, R) \to \ln 2$ as well, independently of the relative fraction size $|F|/|E|$; this independence is sometimes called the Darwinism/objectivity plateau. In Section~\ref{sec:redundancy} below we will characterize more quantitatively the establishment of the objectivity plateau in the QD phaes. Meanwhile, the flow is much slower for $J = 0.45$ (in the intermediate phase, see below). Nevertheless, it is visible that a plateau at a value between $0$ and $1$ establishes itself as $t$ increases. Finally, deep inside the encoding phase, we observe a rapid convergence $\left< r^2\right> \to 0$ regardless of the initial condition. We also observed similar behaviors in the deterministic model. In conclusion, we may indeed treat $k$ as an irrelevant parameter in the more refined study of the phase diagram below. 

\subsubsection{Encoding phase and transition}\label{sec:encoding}
In this section we argue analytically that both models have a stable encoding phase at 
\begin{equation}
J > J_c = \frac12 \,,
\end{equation}
which terminates as a continuous transition at $J_c$ to an intermediate phase.

To show this, we recall that the perfectly encoding distribution $\mathbb{P}(u,v) = \delta(u)\delta(v)$  is always a fixed point of the recursion relation. A model is in the encoding phase if and only if this encoding fixed point is stable (here we assume the uniqueness of the attractive fixed point, see above). In order to study the stability, we may assume that $\mathbb{P}_t$ is a distribution close to the encoding fixed point, and compute second moments of $u$ and $v$ with respect to $ \mathbb{P}_{t+1}$ as a linear map of the same moments with respect to  $\mathbb{P}_{t}$, dismissing higher order terms as being much smaller. 

Concretely, the recursion relation of both models,  \eqref{eq:recursion-deterministic} and \eqref{eq:recursion-random}, implies the following 
\begin{align} 
\left< u^2 \right>_{t+1} =& 
\left<  \frac{(u'_\ell + u'_r)^2}{1 + u'_\ell u'_r} \right>_t \approx  2 \left< (u'_\ell)^2 \right>_t \label{eq:u2order2}  \\ 
\left< v^2 \right>_{t+1} = & 
\left<  \frac{(v'_\ell v'_r)^2}{1 + u'_\ell u'_r} \right>_t\approx  0  \label{eq:v2order2}\\ 
\left< u v \right>_{t+1} = & 
\left<  \frac{(u'_\ell + u'_r) v'_\ell v'_r}{1 + u'_\ell u'_r} \right>_t \approx 0 \,. \label{eq:uvorder2}
\end{align}
Here, the right hand side average is over $\mathbb{P}_t^{\otimes 2}$ and eventually over the random rotation angle; the sign $\approx$ means that the two side are equal up to terms that are higher than second order in $u',v'$; we expand the denominator in geometric series, which is justified as $u',v'$ are small. Note that in \eqref{eq:u2order2} we used the property $\left< u \right> = \left< v \right> = 0$  \eqref{eq:uvmeanis0} to drop $\left< u_{\ell}' u_{r}' \right>$. Now, recalling that  $u'_{\ell} =  \cos\theta_{\ell} u_\ell - \sin\theta_{\ell} v_\ell$, we have 
\begin{equation}
\left< u^2 \right>_{t+1} \approx 2 \cos(J\pi/2)^2 \left< u^2 \right>_{t} + \mathcal{O}(\left< v^2 \right>_t,  \left< uv \right>_t) \,.
\end{equation}
We conclude that the linear recursion map $(\left< (u^2, v^2, uv) \right>_t \mapsto \left< (u^2, v^2, uv) \right>_{t+1}$ has only one nonzero eigenvalue 
\begin{equation}\label{eq:lambda-encoding}
\lambda_c =  2 \cos(J\pi/2)^2 
\end{equation} 
in both models. Thus, when $ 1 > J > J_c = 1/2 $, $|\lambda_c|< 1$, and the encoding fixed point is stable. So we have determined the extent of the encoding phase, as announced above. 

Now, when $J < J_c$, the encoding fixed point becomes unstable. Yet, when $J_c - J$ is small, we may still look for a non-encoding fixed point near the encoding one, by pushing the above calculation to higher order. For this, we let 
\begin{equation}
\epsilon = \cos(J\pi/2)^2 - 1/2 \,,\, 0 < \epsilon \ll 1 
\end{equation}
be the small parameter. We the extend the expansions \eqref{eq:u2order2}-\eqref{eq:uvorder2} further, up to the fourth order in $u',v'$. As a result, we find
\begin{align}
\left< u^2 \right>_{t+1} &  \approx  2 \left< (u'_\ell)^2 \right>_t - 2  \left< (u'_\ell)^2 \right>_t^2   \\ 
\left< v^2 \right>_{t+1}  & 
\approx   \left< (v'_\ell)^2 \right>_t^2 \,,\,  \left< u v \right>_{t+1}  \approx 0 \,. 
\end{align} 
When looking for a fixed point we may assume $ \left< u v \right>_{t} \approx 0$ and $\left< v^2 \right>_t = \mathcal{O} (\left< u^2 \right>_{t-1}^2) \ll \left< u^2 \right>_t $ already. Then we have 
\begin{align}
\left< u^2 \right>_{t+1} &  \approx  (1 + 2\epsilon)  \left< u^2 \right>_{t} + \left< v^2 \right>_t - 2 \left(\frac12 \left< u^2 \right>_t \right)^2   \\ 
\left< v^2 \right>_{t+1} & \approx  \left(\frac12 \left< u^2 \right>_{t} \right)^2  \,,
\end{align}
where we also dropped terms of $\mathcal{O}(\epsilon \left< u^2 \right>^2)$. Plugging the second equation with $t \to t-1$ to the first one, and assuming slow variation $| \left< u^2 \right>_{t}  -  \left< u^2 \right>_{t-1}| \ll \left< u^2 \right>_{t} $, we obtain
\begin{equation}\label{eq:flow_Jc}
\left< u^2 \right>_{t+1}  - \left< u^2 \right>_{t} 
\approx  2 \epsilon \left< u^2 \right>_t - \frac14 \left< u^2\right>_t^2 \,.
\end{equation}
Thanks to the negative quadratic term, this recursion flow has a stable fixed point $\left< u^2 \right>_{t\to\infty} = 8 \epsilon + {o}(\epsilon) \,,\,$
Since $ \left< v^2 \right> $ is much smaller, we have 
\begin{equation}\label{eq:r2-epsilon}
\left< r^2 \right>_{t\to\infty} = 8 \epsilon + {o}(\epsilon) \,.
\end{equation}
We have thus shown that the transition at $J_c$ is a {continuous} one between the encoding phase and an intermediate phase. Note that this argument rules out a direct discontinuous transition into a QD phase. If this were the case, we would not have found a stable fixed point close to the encoding one (assuming uniqueness of stable fixed point). Viewing $r$ as the order parameter of the encoding-intermediate transition, \eqref{eq:r2-epsilon} also predicts the ``spontaneous magnetization'' exponent $\beta$, defined as $  \left< r^2 \right> \sim |J_c - J|^{2\beta} $, to be $1/2$, as in the mean-field theory of magnetism. Section~\ref{sec:coarsegrain} below will provide more substance to this analogy.

\begin{figure}
\centering
\includegraphics[scale=1]{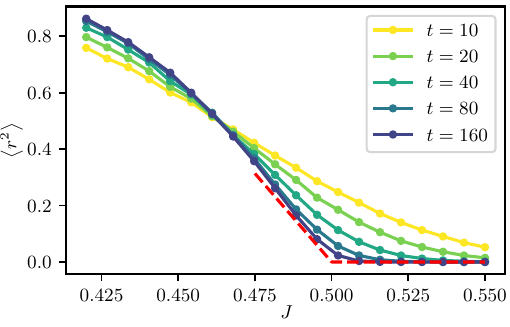}
\caption{The purity $\left< r^2 \right>$ as a function of $J$ near the encoding transition $J_c=1/2$. The numerical data (dots connected by lines, which are a guide to the eye) are obtained in the random model, up to $t = 160$, using the efficient sampling method, see Section~\ref{sec:numerics}. To reduce statistical noise, we averaged over data in a small time interval $[0.9t, t]$ for each data point. The red dashed curve represents the analytical prediction \eqref{eq:r2-epsilon}  in the $t\to\infty$ limit.}
\label{fig:encoding}
\end{figure}
We stress that eq.~\eqref{eq:r2-epsilon} results from a controlled expansion, and is exact, including the pre-factor, in the $t \to \infty$ limit. However, it does not compare quantitatively well with small $t$ numerical data, due to the strong finite size effects near $J \approx J_c$. In fact, in Fig.~\ref{fig:basic} above, it is not even obvious that $J_c = 1/2$. Nevertheless, thanks to the efficient sampling algorithm (see Section~\ref{sec:numerics}) applied to the random model, we are able to go to $t = 160$. The numerical results, plotted in Fig.~\ref{fig:encoding}, show a convincing convergence to the asymptotic prediction~\eqref{eq:r2-epsilon}, with the predicted pre-factor. This agreement is a nontrivial benchmark for the numerical method, which will play an important r\^ole in the analysis of the QD-intermediate transition below. 

\subsubsection{QD phase and transition}\label{sec:QD}
We now turn our attention to smaller values of $J$, where we do not have an exact quantitative theory as we do near the encoding-intermediate transition. For instance, we do not predict the exact locus of the critical point $J_d$ of the QD-intermediate transition. The numerical data from Fig.~\ref{fig:basic} indicate rather convincingly that $J_d > 0$, but are not enough to estimate its location. However, we can make progress by reformulating the problem as the stability of the QD fixed point. 
\begin{figure}
\centering
\includegraphics{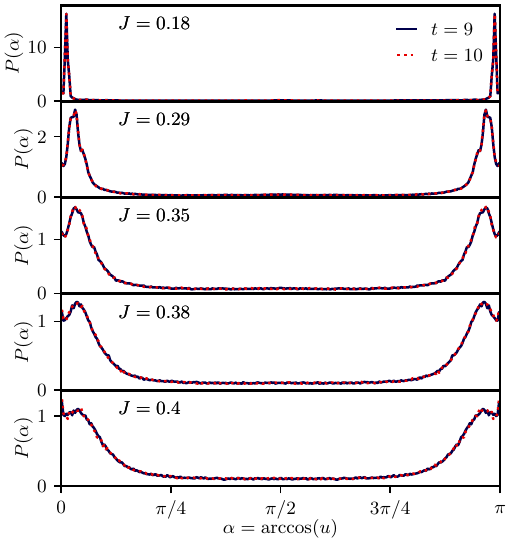}
\caption{Perfectly QD distribution $\mathbb{P}_{t, k=1}$ for a few values of $J$ and $t = 9$ and $t=10$ obtained by solving the recursion relation~\eqref{eq:recursion-random} (random model) with initial condition~\eqref{eq:P0k} and $k = 1$. These distributions are supported on the unit circle $u^2 + v^2 = 1$ and symmetric with respect to $v \to -v$, so we plot them as a distribution of the angle $\alpha = \arccos(u)$. We observe a fast convergence to the $t\to\infty$ limit and no abrupt dependence on $J$. }
\label{fig:Pcircle}
\end{figure}

Indeed, recall that by starting with an initial condition with $k=1$, the distribution $\mathbb{P}_{t,1}$ is perfectly QD for all $t$. Numerically, we find that $ \mathbb{P}_{t,1}$ converges very rapidly to a QD fixed point $\mathbb{P}_{t\to\infty, 1}$; the finite-size effect is much weaker compared to $k > 1$. We also observe that this perfectly QD fixed point depends smoothly on $J$. In Fig.~\ref{fig:Pcircle} we plotted the QD fixed point a few values of $J$. We see that for $J$ small, the distribution is highly concentrated around $u = \pm 1$, and gradually delocalizes as $J$ increases. This indicates that the transition out of the QD phase does not result from a non-analytic $J$-dependence {of the} QD fixed point itself, but rather from its stability with respect to a perturbation that decreases $\left< r^2 \right>$ from $1$. This characterization of the QD-intermediate transition is similar to that of the encoding-intermediate transition, which concerns the stability of the perfectly-encoding fixed point. So, by analogy, we expect the stability can be quantified by some eigenvalue $\lambda_d$, such that  
$$ \left< 1-r^2 \right>_{t+1} = \lambda_d \left<1- r^2 \right>_{t} + o(\left<1- r^2 \right>^2_{t}) \,. $$ 
So, by determining $\lambda_d$ as a functional of the perfectly QD fixed point $\mathbb{P}_{t\to\infty, 1}$, we will be able to leverage the precise numerical estimate of $\mathbb{P}_{t\to\infty, 1}$, and estimate the QD-intermediate transition to be the point where $\lambda_d$ exceeds $1$. 

\begin{figure}
\centering
\includegraphics{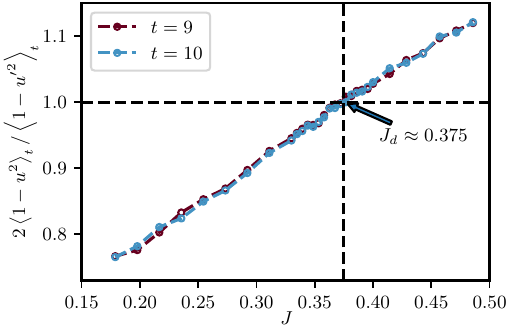}
\caption{The (approximate) stability eigenvalue, $\lambda_d = 2 \left< 1-u^2 \right> / \left< 1 - {u'}^2 \right>$~\eqref{eq:lambdaQD}, evaluated on the perfectly QD distribution $\mathbb{P}_{t, k=1}$ as a function of $J$ for $t = 9, 10$ (the result is practically independent of $t$). See Fig.~\ref{fig:Pcircle} for plots of such distributions. We estimate the QD-intermediate transition $J_d \approx 0.375(5)$ as the value where $\lambda_d = 1$. }
\label{fig:QDstable}
\end{figure}
To carry out this program, we now find an approximate expression of $\lambda_d$. For this let us consider the variable
\begin{equation}
\varrho^2 := \frac{v^2}{1-u^2} \,.
\end{equation}
Its average $\left<  \varrho^2 \right> = 1$ for a perfectly QD distribution, just like $\left< r^2 \right>$. Moreover, the recursion relation~\eqref{eq:recursion-random} implies the following exact identity:
\begin{align}
2   \left<  \varrho^2 \right>_{t+1} &=  \left< \frac{v^2}{1-u} \right>_{t+1} + \left<  \frac{v^2}{1 + u} \right>_{t+1}  \nonumber \\ 
&=  \left< \frac{{v'}^2}{1-{u'}} \right>^2_t +  \left< \frac{{v'}^2}{1 + {u'}} \right>^2_{t} \nonumber \\ 
&= \left< {\varrho'}^2 (1+u') \right>^2 + \left<  {\varrho'}^2 (1 - u') \right>^2 \,,
\end{align}
where we dropped the $\ell,r$ subscript from $u', v'$ and wrote $\varrho'^2 = {v'}^2 / (1- {u'}^2)$. Now, let us assume that $\mathbb{P}_t$ is close to the QD fixed point, so in particular $\varrho'$ is close to $1$, and expand the right hand side up to linear order in $1-{\varrho'}^2$. Recalling $\left< u' \right> = 0$, we find a nice formula:
\begin{equation}
\left<  1- \varrho^2 \right>_{t+1} = 2 \left<  1 - {\varrho'}^2 \right>_t + \mathcal{O}((1-\varrho')^2) \,. \label{eq:rho2times}
\end{equation}
It remains to relate $\langle 1- {\varrho'}^2 \rangle$ with $\left< 1 - {\varrho}^2\right>$. To do this we observe that since $(u',v')$ is $(u,v)$ rotated by $\theta$, $r^2 = {r'}^2$. Moreover, we also have $1-r^2 = (1-\varrho^2) (1-u^2)$. To proceed further, we make the following uncontrolled approximation:
\begin{equation}\label{eq:approximation}
\left< 1-r^2 \right>  = \left<  (1-\varrho^2) (1-u^2) \right> \approx \left<  1-\varrho^2 \right>  \left< 1-u^2 \right> \,,
\end{equation}
and similarly for $r', \varrho', u'$. Essentially, we consider the amplitude of the perturbations $ (1- {\varrho'}^2)$, $ (1-\varrho^2)$ to be uniform, and thus independent of $1-u^2$. Then we obtain 
$$ \left<  1 - {\varrho'}^2 \right> \approx \frac{ \left< 1-{r'}^2 \right>}{ \left< 1- {u'}^2 \right>} = 
\frac{ \left< 1-{r}^2 \right>}{ \left< 1- {u'}^2 \right>} \approx \left<  1-\varrho^2 \right>\frac{ \left< 1-u^2 \right> }{ \left< 1- {u'}^2 \right>} \,. $$
Combining this with \eqref{eq:rho2times} we conclude that 
\begin{align} \label{eq:lambdaQD}
&   \left<  1- \varrho^2 \right>_{t+1} \approx \lambda_d \left<  1 - {\varrho'}^2 \right>_t, \,  \lambda_d := 2 \frac{\left< 1-u^2 \right>_t }{ \left< 1- {u'}^2 \right>_t} \,.
\end{align}
This is the stability ``eigenvalue'' of the QD fixed point we sought for.  If $\lambda_d > 1$ ($\lambda_d < 1$), a small perturbation away from the manifold of perfectly QD distribution will be amplified (shrunk, respectively), and thus the QD fixed point is unstable (stable). Qualitatively, it indicates that a QD fixed point tends to be more stable if it is peaked around $u = \pm 1, v = 0$ (since then the random rotation will make $1- {u'}^2$ larger). This is after all expected: the branching isometry $Y$~\eqref{eq:Ydef} broadcasts information only in the $z$ direction. 


Quantitatively, eq.~\eqref{eq:lambdaQD} is useful as an efficient approximate numerical method to test the stability of QD fixed points, and thus locate the critical point $J_d$. As we only need to evaluate the expression of $\lambda_d$ on the fixed point, we avoid the strong finite-size effects exhibited by the flow of intermediate distributions (see Fig.~\ref{fig:basic}). We applied this method to the numerically converged QD fixed points for a range of $J$. The result is plotted in Fig.~\ref{fig:QDstable}. We find that $\lambda_d$ increases with $J$, and exceeds $1$ at 
\begin{equation} \label{eq:Jd}
J = J_d \approx 0.375(5) \quad \text{(random model)}
\end{equation} 

\begin{figure}
\centering
\includegraphics{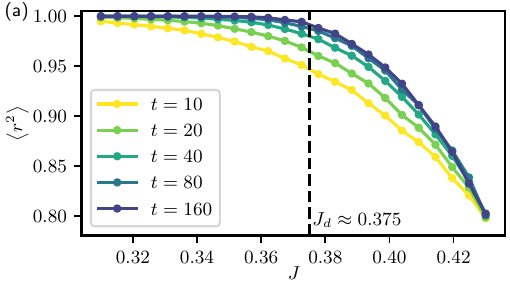}
\includegraphics{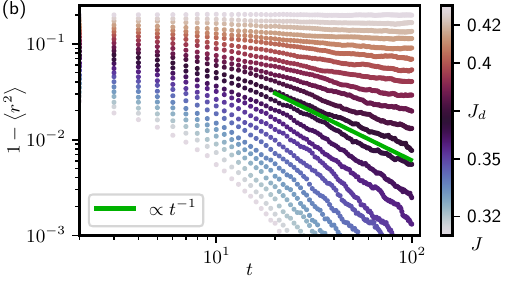}
\caption{Purity $\left< r^2 \right>$ in the random model up to $t =160$, obtained by numerical solution to the recursion relation~\eqref{eq:recursion-random}, see Section~\ref{sec:numerics} for numerical methods. (a) $ \left< r^2 \right>$ as a function of $J$, for $t =10, 20, \dots, 160$. To reduce statistical noise, we averaged over data in a small time interval $[0.9t, t]$ for each data point. The vertical dashed line represents the estimate of the QD-intermediate critical point \eqref{eq:Jd}. (b) $ 1-\left<r^2 \right>_t $ as a function of $t$, for different values of $J$ (see color bar). The green line indicates a power law decay $\propto t^{-1}$. The data curves above (below) the green line have $J > J_d$ ($J < J_d$), respectively.}
\label{fig:longtime}
\end{figure}
To test this estimate, and further characterize the critical behavior, we performed extensive simulations up to $t = 160$. This is feasible thanks to efficient sampling method presented in Section~\ref{sec:numerics}, and bench-marked in Section~\ref{sec:encoding} in the slow, $J \approx 0.5$ regime. We first plot the results as a function of $J$ for increasing $t$, in Fig.~\ref{fig:longtime}~(a). The results are in nice agreement with \eqref{eq:Jd}, and also suggest a ``mean-field'' behavior of the ``order parameter''  $ \phi_t := 1 - \left< r^2 \right>_t $, namely, $\phi_{t\to\infty} \sim |J - J_d|$ at $J > J_d$. 

To corroborate this claim, we plot in Fig.~\ref{fig:longtime}-(b) the same data as a function of $t$. We find that near criticality, $m_t \sim t^{-1}$ decays as a power law. This power law is a signature of the following ``mean-field'' effective flow equation at small $t$ [compare to \eqref{eq:flow_Jc}]:
\begin{equation} \label{eq:MFflow}
\partial_t \phi_t =  (\lambda_d - 1) \phi_t - b \phi_t^2 + o(\phi_t^2) \,,\, b > 0.
\end{equation}
Here, the quadratic term has a negative sign and hence the linear term controls the stability of the QD fixed point $\phi_t$, and  the coefficient must be $\lambda_d - 1$, by definition of $\lambda_d$. At criticality, $\lambda_d = 1$ and we have $\partial_t \phi_t \propto - \phi_t^2$ and hence $\phi_t \sim t^{-1}$, which is observed in Fig.~\ref{fig:longtime} (b). When $\lambda_d < 1$ the power law decay crosses over to an exponential one eventually. When $\lambda_d > 1$, the decay halts at the stable fixed point $\phi_{t\to\infty} \propto (\lambda_d - 1) $. Since $\lambda_d$ depends smoothly on $J$ such that $\lambda_d - 1 \propto J - J_d$ (see Fig.~\ref{fig:QDstable}), we have indeed $\phi_{t\to\infty} \sim (J - J_d)$.

\begin{figure}
\centering
\includegraphics{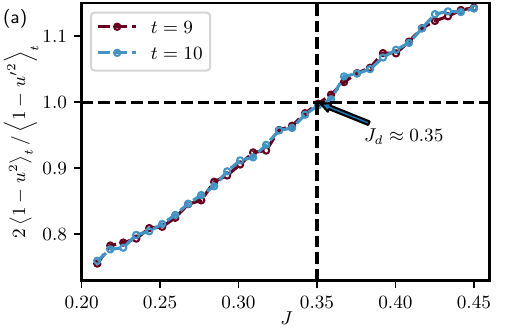}
\includegraphics{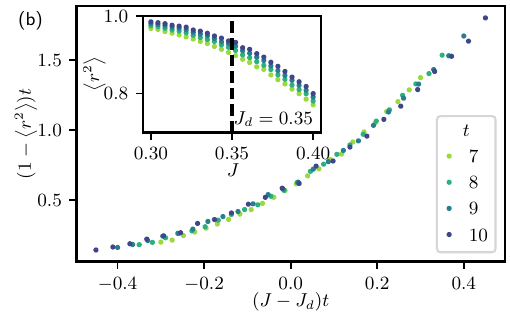}
\caption{Numerical study of the QD-intermediate transition in the deterministic model. (a) The approximate stability eigenvalue $\lambda_d = 2 \left< 1 - u^2\right> / \left< 1 -{u'}^2\right>$~\eqref{eq:lambdaQD} as a function of $J$, evaluated on the perfect QD distribution $\mathbb{P}_{t,k=1}$ with $t = 9$ and $10$, at which value $\mathbb{P}_{t,k=1}$ has well converged to the $t = \infty$ limit. The QD-intermediate critical point $J_d \approx 0.35(1)$ is estimated as the value where $\lambda $ exceeds $1$. (b) Main: Testing the scaling Ansatz \eqref{eq:ansatz} with $J_d = 0.35$ (not adjusted) using numerical data from $t = 7, \dots, 10$ ($k = 2$). Inset: raw data. } 
\label{fig:QD-det}
\end{figure}
So far we focused on the random model where extensive simulation is possible. Concerning the deterministic model, the approximate method based on estimating the stability of the QD fixed point is still applicable. Indeed, we observe a fast convergence of the recursion relation flow with the $k=1$ initial condition as in the random model. As a result, we obtain an estimate:
\begin{equation}
J_d \approx 0.35(1) \quad \text{(deterministic)} \,,
\end{equation}
see Fig~\ref{fig:QD-det} (a). To test this estimate, we computed $\mathbb{P}_{t, k=2}$ up to $t = 10$ and tested the purity against the scaling Ansatz 
\begin{equation}\label{eq:ansatz}
( 1-\left< r^2 \right>_t)  t = f( (J-J_d)t)  \,,
\end{equation}
which can be derived from the mean-field effective flow equation~\eqref{eq:MFflow}. As a result, we observe a reasonable collapse of data with different $t$ in Fig.~\ref{fig:QD-det}-(b), with \textit{no} adjustable parameters. These results indicate that in both models there is a QD phase and a QD-intermediate transition of similar mean-field critical properties. 

\subsubsection{Scaling of redundancy}\label{sec:redundancy}
Let us now apply the tools developed so far and revisit the fraction size dependence, making connection with the notion of  \textit{redundancy}~\cite{zurek-QD,zurek-review}. In Quantum Darwinism, this notion is defined as the inverse of the minimal relative fraction size such that the Holevo bound with the reference is greater than $(1-\delta)$ bit, where $\delta$ is the tolerance: 
\begin{equation}\label{eq:Rdelta-def}
R_\delta =  1 /  \min \{|F| / |E|: \chi(F, R) \ge  (1-\delta) \ln 2 \} \,. 
\end{equation} 
In other words, $R_{\delta}$ estimates the number of ``good'' copies of the injected information that are broadcast into the environment; the smaller $\delta$ is, the stricter the standard of ``good'' is. Since the quantum discord vanishes for the microscopic measurement, the result below applies \textit{verbatim} to the mutual information $I(F,R) = \chi(F,R)$.

We would like to understand how $R_\delta$ scales with $\delta$ and $t$ in the QD phase. For this, consider a small relative fraction size 
\begin{equation}
|F| / |E| = 2^{-k} \,,\, R = 2^{k} \,,\, k \gg 1\,.
\end{equation}
This leads to an almost encoding initial condition~\eqref{eq:P0k}, such that 
\begin{equation}
\left<  u^2 \right>_{t=0} \sim c^{2k} \ll 1 \,,\, c := \cos(J \pi /2 )\,,\, 
\end{equation}
Then, initially, we may apply the linearized recursion relation at the encoding fixed point and find~\eqref{eq:lambda-encoding}
\begin{equation}
\left<  u^2 \right>_{t} \sim  c^{2k} \lambda_c^{t} \,,\, \lambda_c = 2 c^2 > 1 \,.
\end{equation}
The above equation holds as long as the RHS is small, that is,  \begin{equation}
t  \lesssim t_k =  2 k \, \frac{ |\ln c|}{ \ln \lambda_c }  \,.
\end{equation}
Then, after a transient of order one duration, the flow will converge to a QD fixed point exponentially with a rate given by the stability eigenvalue of the QD fixed point $\lambda_d < 1$ [an approximate formula of $\lambda_d$ is given by \eqref{eq:lambdaQD}]:
\begin{equation}
\phi :=  \left< 1 - r^2 \right> \sim \lambda_d^{t - t_k} \,. 
\end{equation}
Now recall from~\eqref{eq:chiFR} that $\phi$ is essentially the distance between the Holevo bound and $\ln2$, up to a log-correction:
\begin{equation}
\delta = 1 - \frac{I(F,R)}{\ln2} \sim {\phi |\ln \phi|}  \implies \phi \sim \frac{\delta}{|\ln \delta|} \,.
\end{equation}
Combining the above, and letting $ n = t + k$ (recall that $|E| = 2^n$ is the total size of the environment), we obtain the scaling law of redundancy:
\begin{equation} \label{eq:Rdelta-scaline}
R_\delta \sim |E|^{\frac{\ln \lambda_c}{\ln 2}}   \left(\frac{\delta}{|\ln \delta|}\right)^{ \frac{\ln \lambda_c}{|\ln \lambda_d|}}  \,.
\end{equation}
A few observations are in order. First, the redundancy grows as a power law of the environment size. The exponent $\ln \lambda_c / \ln 2 < 1$, approaching $1$ from below as $J \to 0$. Thus, in the thermodynamic limit, the redundancy tends to infinity for any fixed tolerance, yet remains \textit{sub-extensive}: the number of redundant copies is much smaller than that of the degrees of freedom. 

It is worth remarking that the exponent ${\frac{\ln \lambda_c}{\ln 2}} $ solely depends on the instability eigenvalue of the \textit{encoding} fixed point. In fact, the $|E|$ scaling of \eqref{eq:Rdelta-scaline} applies also in the intermediate phase, as long as the tolerance is not too small so that $(1-\delta) \ln 2 < \lim_{t\to\infty} \chi(F, R)$: we have a redundant yet imperfect broadcast of the injected information. 

In the QD phase, $R_\delta$ decays as a power law of the tolerance $\delta$. The exponent depends on the stability eigenvalues of both QD and encoding fixed points. It diverges at the QD-intermediate transition (since $\lambda_d \to 1$). Thus, when $J \sim J_c$, large good quality redundancy is almost impossible to achieve. In fact, redundancy remains small in moderate-sized systems even away from $J_c$. For example, take $J = 0.2$, for which $\lambda_d \approx 0.75$ (in both models). Then according to \eqref{eq:Rdelta-scaline}, $R_{10\%} \approx 0.5$ and $R_{20\%} \approx 5$ for $|E| = 10^3$, while $R_{10\%} \approx 30$ and $R_{20\%} \approx 256$ for $|E| = 10^5$. This is after all unsurprising, as the models we are considering are by design near the boundary of the QD phase. 

\subsubsection{Summary}
We studied the information retrieval phase diagram of the tree models. We found that there are three phases, the QD phase at $ 0 < J < J_d$, the intermediate phase $J_d <J<J_c$ and the encoding phase $ J_c < J < 1$. The phase diagram is independent of the relative size of the fraction. The encoding-intermediate transition takes place at $J_c = 1/2$ (exact) for both models. The QD-intermediate critical point is numerically estimated as $J_d \approx 0.375(5)$ in the random model and $J_d \approx 0.35(1)$ in the deterministic one. Both transitions display mean-field critical behavior. 

Note that the above phase diagram is specific to the microscopic measurements, where we measure all the spins in $F$ in the computational basis. Since the quantum discord vanishes for this measurement, the phase diagram also describes the mutual information $I(F,R)$, and thus the maximal amount of injected information one may retrieve. $I(F,R) \to \ln2$ in the QD phase, $I(F,R) \to 0$ in the encoding phase, and in the intermediate phase, the $I(F,R)$ tends to a $J$-dependent value between $0$ and $\ln2$. 

\subsection{Coarse-grained measurements}\label{sec:coarsegrain}
In this section we study the information retrievable by coarse-grained measurements. One motivation is to address the relation between measurement result and retrieved information. In general, this relation is given by the mapping $ m \mapsto Q_{m}$. In a tree model, this can be computed numerically without much difficulty for any given microscopic outcome $ m = (m_i)_{i\in F}$. However, to understand the relation in the thermodynamic limit, one clearly needs a coarse-grained description of the outcome. To make an analogy with standard equilibrium statistical mechanics, $ Q_{\vec{m}}$ corresponds to the Boltzmann weight of a microscopic configuration, $e^{-H[\sigma]}$ (in this sense, $ Q_{\vec{m}}$ is an \textit{operator-valued} Boltzmann weight). In statistical physics, it is well-known that by coarse-graining the microscopic configuration $\sigma \to \varphi$, we may obtain an effective action $e^{-S_{\text{eff}} (\varphi)}$ of the collective field, which is more useful in describing emergent behaviors. The results below can be viewed as a first glimpse of the (operator-valued) effective action in our context.   

Throughout this section we shall focus on the deterministic model. In Section~\ref{sec:coarse-grain1} we derive the recursion relation for the total spin measurement and present numerical results. In Section~\ref{sec:noQD} we will focus on the small $J$ regime, and show that the QD phase cannot be observed with coarse-graining measurements. Meanwhile, they are able to probe the encoding-intermediate transition, as we will show in Section~\ref{sec:encoding-coarse}. 

\subsubsection{Recursion relation and numerical results}\label{sec:coarse-grain1}
We start with the most basic coarse-grained measurement, that of the total spin of the fraction $F$. We can think of it as measuring all the spins in $F$ as above, but then coarse-graining the outcome $\vec{m} = (m_i)_{i\in F}$ to one integer, their sum:
\begin{equation} \label{eq:Mtotaldef}
\mathcal{M} = \sum_{i \in F} m_i \,.
\end{equation}
Hence, the corresponding projectors are 
$$\pi_{\mathcal{M}} = \sum_{\sum m_i = \mathcal{M}} \pi_{\vec{m}} .$$ The \textit{non-normalized} post-measurement density matrices $Q_{\mathcal{M}} = V^\dagger \pi_{\mathcal{M}} V$ \eqref{eq:Qmdef} of the coarse-grained measurements can be obtained from the microscopic measurement:
\begin{equation}
Q_{\mathcal{M}} = \sum_{\substack{ 
		\sum_i m_i = \mathcal{M}}} Q_{\vec{m}}  \,.
		\end{equation}
		We will parametrize them in a slightly different way from \eqref{eq:defparamuv} above, as follows:
		\begin{align}\label{eq:QMuv}
Q_{\mathcal{M}} =& p_{\mathcal{M}} (\mathbf{1} +  u_{\mathcal{M}} \sigma^z  + v_{\mathcal{M}} \sigma^x)  \nonumber \\ 
=& p_{\mathcal{M}} \mathbf{1} +  a_{\mathcal{M}} \sigma^z  + b_{\mathcal{M}} \sigma^x  \,. 
\end{align}
That is, we write $a = p u$ and $b = p v$, which will prove convenient. Recall that $p_{\mathcal{M}} $ is the probability of the outcome $\mathcal{M}$, while $(\mathbf{1} + u_{\mathcal{M}}\sigma^z  + v_{\mathcal{M}}\sigma^x )/2$ is the normalized post-measurement density matrix of the reference. So we can define a outcome-resolved purity:
\begin{equation}
r^2_{\mathcal{M}} := u^2_{\mathcal{M}} + v^2_{\mathcal{M}} \,.
\end{equation} 
Further averaging over the outcomes gives us back the ensemble-averaged purity: 
\begin{equation}
\left< r^2 \right> = \sum_{\mathcal{M}} r^2_{\mathcal{M}} p_\mathcal{M} \,.
\end{equation}
Note that the left hand side is in general smaller than the same quantity with microscopic measurements: by using less of the measurement outcome, we retrieve less information about the reference. Of course, $p_{\vec{m}}, a_{\vec{m}}, b_{\vec{m}}$ are defined for the microscopic measurement as well. However, coarse-graining makes it feasible to calculate such quantities in large systems. 

\begin{figure}
\centering
\includegraphics{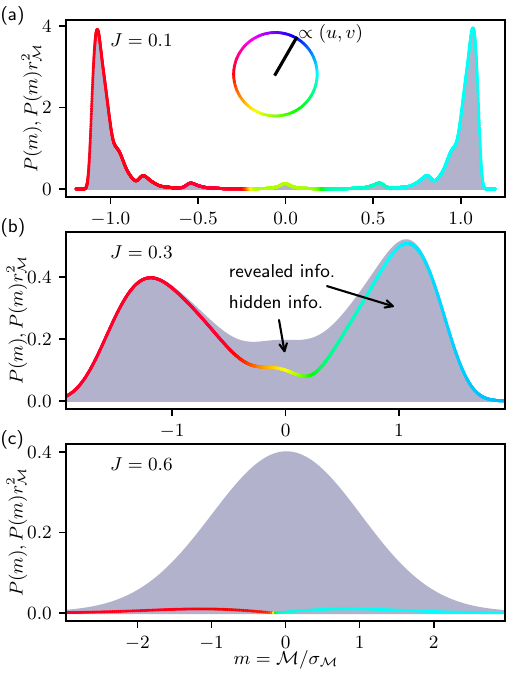}
\caption{Outcome distribution of $\mathcal{M} = \sum_{i \in F} m_i$ (total spin of the fraction) and the post-measurement reference state, for $J =0.1$ (a), $0.3$ (b) and $0.6$ (c) in the deterministic model with fraction size $|F| = 2^{12}$ and relative size $|F|/|E| = 1/4$. (See also Fig.~\ref{fig:peaks} for a closer look at very small $J$.) We plot the probability density of the total spin re-scaled by its standard deviation, $m = \mathcal{M} / \sigma_{\mathcal{M}}$, $P(m) = p_{\mathcal{M}} \sigma_{\mathcal{M}}$ as the filled area. $r^2_{\mathcal{M}} P(z)$ is plotted as a colored curve, where the color indicates the angle of $(u_{\mathcal{M}},v_{\mathcal{M}})$, indicating the polarization direction the post-measurement reference state. The filled area below and above the colored curve represents the information revealed by and hidden from the measurement respectively.} 
\label{fig:PMs}
\end{figure}
To proceed, we derive a backward recursion relation for $ Q_{t,\mathcal{M}}$, where we recall that $t$ indicates the size of the fraction $|F| = 2^t$. For this, we observe that $\mathcal{M} = \mathcal{M}_\ell + \mathcal{M}_r$ where $\mathcal{M}_\ell$ and $\mathcal{M}_r$ are the total magnetization of the left and right sub-tree of the root, respectively. Then, using the recursion relation for $Q_{\vec{m}}$ \eqref{eq:Qrecursion}, we have
\begin{equation}\label{eq:recursion-totalspin-gen}
Q_{t+1,\mathcal{M}} = \sum_{\mathcal{M}_\ell + \mathcal{M}_r=\mathcal{M}}  \mathcal{L}_{\hat{Y}} \left( Q_{t,\mathcal{M}_\ell} \otimes Q_{t,\mathcal{M}_r} \right)
\end{equation} 
This holds for a general deterministic binary tree model. To write the recursion relation for our concrete deterministic model, we shall use parametrization \eqref{eq:QMuv}, and the shorthand for rotation 
\begin{equation}\label{eq:rotation-ab}
\begin{pmatrix}
	a' \\ b'
\end{pmatrix} = \begin{pmatrix}
	\cos \theta & -\sin \theta \\
	\sin \theta & \cos \theta  
\end{pmatrix} \begin{pmatrix}
	a \\ b
\end{pmatrix} \,,\, \theta = J \pi / 2 \,.
\end{equation}
It is also convenient to introduce the discrete convolution with respect to the variable $\mathcal{M}$:
\begin{equation}
(f \star g)_{\mathcal{M}} =  \sum_{\mathcal{M}_\ell + \mathcal{M}_r=\mathcal{M}} f_{\mathcal{M}_\ell} g_{\mathcal{M}_r} \,. 
\end{equation}
Then, it is not hard to obtain the following recursion relations
\begin{align}
& p_{t+1} = p_t \star p_t +  a'_t \star a'_t, \nonumber \\
& a_{t+1}  = 2 p_t \star a'_t,   \label{eq:recursion-coarse}\\ & b_{t+1}  =  b'_t \star  b'_t.\nonumber
\end{align}
The initial condition is the same as for the microscopic measurement, since $\mathcal{M} = m$ when $F$ has only one spin, and we recall it here~\eqref{eq:Qminit}:
\begin{equation}\label{eq:init-coarse}
p_{0,\mathcal{M}} = \frac12 \,,\, a_{0,\mathcal{M}} = \frac12  \mathcal{M} c^{k-1} \,,\, b_{0, \mathcal{M}} = 0  \,,
\end{equation}
where $\mathcal{M} = \pm 1$, $|F|/|E| = 2^{-k}$ and $c = \cos (J\pi/2)$. Note that $p_t, a_t, a_t$ are all functions of the discrete variable $\mathcal{M}$ that can take $2^t+1$ values. Hence, computing exactly $p_t, a_t, b_t$ using the above recursion relations is only exponentially hard in $t$, or linearly hard in the fraction size $|F|$. This is far better than microscopic measurement case, where the brute-force computation cost is $\mathcal{O}(2^{|F|})$.  Moreover, in practice $Q_{\mathcal{M}}$ depends smoothly on $\mathcal{M}$ and can nearly vanish for most $\mathcal{M}$, so that we can represent the function $Q_{\mathcal{M}}$ in a compressed way and still obtain numerically exact results.   

In Fig.~\ref{fig:PMs} we plot the exact numerical results thus obtained for $t = 12$, for three values of $J$. We observe that, for a large $J = 0.6$, the outcome distribution resembles a Gaussian with zero mean. Also, $r^2_{\mathcal{M}}$ almost vanishes for any $\mathcal{M}$. This is expected since $J > J_c = 1/2$ is in the encoding phase where $F$ and $R$ are uncorrelated in the thermodynamic limit. As $J$ decreases below $1/2$, the distribution of $\mathcal{M}$ deviates from being Gaussian, and develops two peaks at positive and negative values. Moreover, $\mathcal{M}$ becomes correlated with the $z$-component of the reference. For $\mathcal{M}$ on the positive (negative) peak, the references post-measurement state $\rho_{\mathcal{M}}$ is close to $| + \rangle_z$ ($| - \rangle_z$, respectively), so that $r^2_{\mathcal{M}} \to 1$. Meanwhile, $0<r^2_{\mathcal{M}}<1$ when $\mathcal{M}$ is between the peaks: obtaining such an outcome reveals less information about the reference. Finally, as $J$ approaches $0$, the distribution has two sharp and separated peaks. $r^2_{\mathcal{M}}$ becomes close to $1$ for all values of $\mathcal{M}$, including when $\mathcal{M} \approx 0$. For these values, $\rho_{\mathcal{M}} \approx | - \rangle_x$ is polarized in the $x$ direction. 

Globally, the outcome distribution $p_{\mathcal{M}}$ is qualitatively reminiscent of that of the total magnetization in the Ising model, which is also a Gaussian in the paramagnetic phase and non-Gaussian with two peaks in the ferromagnetic one. Meanwhile, $u_{\mathcal{M}}$ and $ v_{\mathcal{M}})$ have no analogy with classical magnetism, and are the new ingredients of the operator-valued effective action.

\subsubsection{Absence of the QD phase}\label{sec:noQD}
\begin{figure}
\centering
\includegraphics{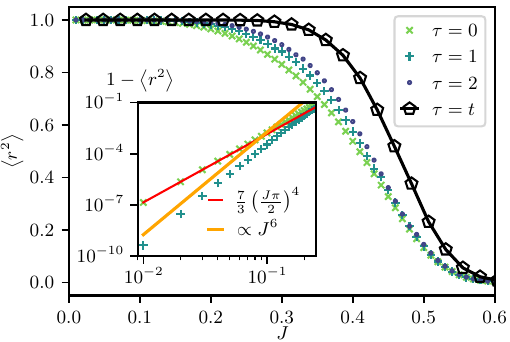}
\caption{Main: the averaged purity with different measurement schemes in the deterministic model with fraction size $|F| = 2^{12}$ and relative size $|F| / |E|= 2^{-2}$. $\tau = 0$: only the total spin of $F$ is measured. $\tau = 1$: the total spins of the left and right subtrees are measured separately. $\tau = 2$: the total spins of the four ``grand-children'' subtrees of the root are measured separately. $\tau = t$: the microscopic measurement. See also Fig.~\ref{fig:coarsegraintree}. Inset: log-log plot of the $\tau = 1$ and $\tau = 2$ data near $J = 0$, compared to power laws. The prefactor of the $J^4$ power law is an exact prediction~\eqref{eq:J4prefactor}. }
\label{fig:sumZ}
\end{figure}
Let us now be more quantitative, and average the purity over the measurement outcomes, and compare the results to the microscopic measurements. In Fig~\ref{fig:sumZ}, we see that coarse-graining decreases the purity, as expected. While the difference is quantitatively more apparent for intermediate values of $J$, there is a qualitative change at small $J$: under coarse-graining, the purity $\left< r^2 \right>$ does not tend to $1$ in the thermodynamic limit for any nonzero $J$. Instead, we have
\begin{equation}  \label{eq:J4}
1- \left< r^2 \right>_{t\to\infty} \sim J^{4} \,,\, J \to 0 \,.
\end{equation}
In other words, the nontrivial QD phase disappears with the total spin measurement. When $J$ is small the information retrieval is almost perfect for all practical purposes, but not strictly perfect in the thermodynamic limit. 

\begin{figure}
\centering
\includegraphics{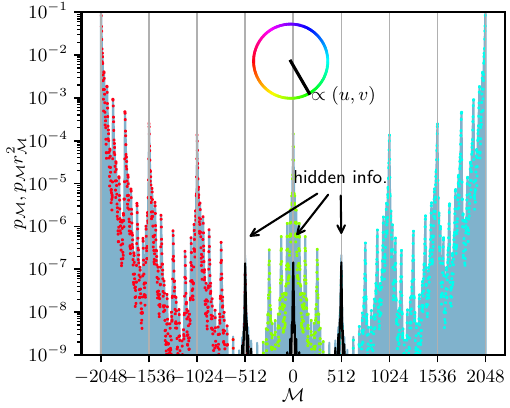}
\caption{Outcome distribution of the total spin measurement with $J = 0.02$, $|F|=2^{11}$, $|F|/|E|=2^{-2}$. See Fig.~\ref{fig:PMs} for further description. Here, in addition, we plot $p_{\mathcal{M}} (1 - r_{\mathcal{M}}^2)$ in black. At small $J$, the distribution has a ``multi-fractal'' structure of peaks; the dominant ones are $\mathcal{M} \approx \pm |F|$. The information retrieval is nearly perfect: $r^2_{\mathcal{M}}$ is close to $1$ for all $\mathcal{M}$. The imperfections are concentrated at secondary peaks at $\mathcal{M} = 0, \pm |F| / 4$. }
\label{fig:peaks}
\end{figure}
The basic reason behind the absence of a nontrivial QD phase is that the total-spin recursion relation~\eqref{eq:recursion-totalspin-gen} does \textit{not} send a perfectly QD distribution to another one, in contrast with the microscopic measurement case~\eqref{eq:r2fixedpoints}. Indeed, suppose that in the RHS of \eqref{eq:recursion-totalspin-gen}, $Q_{t,\mathcal{M}}$ is always proportional to a pure state, and thus so is each single term in~\eqref{eq:recursion-totalspin-gen}. But their sum will be mixed, unless all the terms are proportional to each other. To see how this works in our model,  we followed the recursion relation~\eqref{eq:recursion-coarse} for generic $\theta$ (using symbolic algebra software), starting from a perfectly QD initial condition: \eqref{eq:init-coarse} with $k = 1$. We found that the $t = 1$ distribution is still perfectly QD; at $t = 2$, the first imperfection appears at $\mathcal{M} = 0$, where $p_{\mathcal{M}} (1-r^2_{\mathcal{M}}) = \theta^4 + \mathcal{O}(\theta^5)$. Starting from $t = 3$, we find consistently that the leading imperfections are 
\begin{equation} \label{eq:deficit}
p_{\mathcal{M}} (1-r^2_{\mathcal{M}}) = \begin{cases}
	\theta^2 + \mathcal{O}(\theta^5) & \mathcal{M} = 0 \\
	2 \theta^2 / 3 + \mathcal{O}(\theta^5) & \mathcal{M} \pm =  2^{t-2}  \\
	\mathcal{O}(\theta^5) & \text{otherwise} 
\end{cases} \,.
\end{equation}
So we conjecture that the prefactor in \eqref{eq:J4} is given by
\begin{equation}\label{eq:J4prefactor}
\left< 1-r^2 \right> \sim \frac{7}4 \left(\frac{ J\pi}2\right)^4 \,,\, J \to 0 \,.
\end{equation} 
at $t\to\infty$. This prediction is corroborated by (finite-precision) numerics, see Fig.~\ref{fig:sumZ} (inset). The structure of the leading imperfection is shown in Fig.~\ref{fig:peaks}, which also reveals a beautiful ``multi-fractal'' structure of $p_{\mathcal{M}}$ at small $J$. The distribution appears singular, made of peaks whose amplitudes differ by orders of magnitudes. It would be interesting to characterize this structure with a systematic small-$J$ expansion.  

\begin{figure}
\centering
\includegraphics[width=.8\columnwidth]{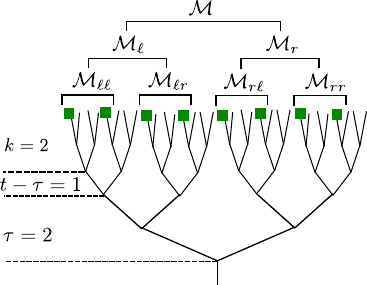}
\caption{Illustration of coarse-grained measurements. The green squares indicate the spins in the fraction $F$, with size $|F| = 2^{t} = 8$ and relative size $|F| / |E| = 2^{-k} = 1/4$. $\mathcal{M}$, the total spin of $F$, is divided into that of the subtrees. The coarse-grained measurement with $\tau = 2$ will measure $\mathcal{M}_{\ell\ell}, \dots, \mathcal{M}_{rr}$. The corresponding density matrices can be obtained in two stages. First we obtain the total-spin ($\tau=0$) matrices at $t - \tau = 1$ using the recursion~\eqref{eq:recursion-coarse}; then we iterate~\eqref{eq:recursion-tau} twice ($\tau = 2$). }
\label{fig:coarsegraintree}
\end{figure}

The total spin measurement is the first and simplest one of a sequence of coarse-grained measurements. For each $\tau \ge 0$, we can consider measuring the total spin of the $2^\tau$ sub-trees of depth $\tau$ from the root. The case $\tau = 0$ is the total spin measurement considered above. For $\tau = 1$, we measure the total spin of the two direct descendant sub-trees of the root, $(\mathcal{M}_\ell,\mathcal{M}_r)$. For  $\tau = 2$ we measure that of the four ``grandchildren'' sub-trees of the root, $(\mathcal{M}_{\ell \ell},\mathcal{M}_{\ell r},\mathcal{M}_{r \ell},   \mathcal{M}_{r r})$, and so on. (In all cases, only the spins in the fraction $F$ are included.) The microscopic measurement corresponds to $\tau = t$, since then each sub-tree has only one spin in $F$. See Fig.~\ref{fig:coarsegraintree} for an illustration. It is not hard to show that the non-normalized density matrices of all such coarse-grained measurements can be obtained recursively:
\begin{align} \label{eq:recursion-tau}
&    Q^{t+1}_{\mathcal{M}_\ell, \mathcal{M}_r} = \mathcal{L}_{\hat{Y}}( Q^t_{\mathcal{M}_\ell}\otimes Q_{\mathcal{M}_r}) \,, \\ 
&     Q^{t+1}_{\mathcal{M}_{\ell \ell},\mathcal{M}_{\ell r},\mathcal{M}_{r \ell},   \mathcal{M}_{r r}} = \mathcal{L}_{\hat{Y}}( Q^t_{\mathcal{M}_{\ell \ell},\mathcal{M}_{\ell r} }\otimes Q^t_{\mathcal{M}_{r \ell},\mathcal{M}_{r r} } ) \,, \nonumber
\end{align}
and so on. (We indicate $t$ and $t+1$ as superscripts instead of subscripts for display.) Note that the recursion increases both $t$ and $\tau$. In fact, these recursion relations are essentially identical to \eqref{eq:Qrecursion}, except that the initial condition is given by $Q_{t,\mathcal{M}}$ with $t \ge 0$. Direct numerical calculation using the above recursion has a cost $2^{(t-\tau) 2^\tau}$. In practice, thanks to compression (see Section~\ref{sec:coarse-grain1}), we can easily obtain reliable results up to $\tau = 2$ and $t = 12$. The results are plotted in Fig.~\ref{fig:sumZ} as well. As $\tau$ increases, we retrieve more information about the reference. However, a nontrivial QD phase cannot be recovered with any finite $\tau$, since applying the recursion map a finite times cannot turn an intermediate distribution to a perfectly QD one. Numerically we find that for small $J$ $1 - \left< r^2 \right>_t \sim J^{a(\tau)} $ where the exponent increases with $\tau$, with $a(0) = 4$~\eqref{eq:J4}, $a(1) \approx 6$, etc. In the fine-grained ($\tau \to \infty$) limit, the distinction between QD and intermediate phases re-emerges. Note that we can obtain the fine-grained limit by sending $t\to\infty$ before $\tau$, so $1 \ll \tau \ll t$. In other words, to probe the QD phase requires arbitrarily fine resolution, but not necessarily a microscopic one.  

\subsubsection{Encoding-intermediate transition}\label{sec:encoding-coarse}
We now come back to $\tau = 0$ (total spin measurement) and study the recursion relations \eqref{eq:recursion-coarse} analytically. We will show that the total spin measurement exhibits an encoding-intermediate transition at $J = 1/2$, which coincides with the same transition under microscopic measurements. We will also show that the outcome distribution tends to a Gaussian in the encoding phase, and quantify the non-Gaussianity when $J < 1/2$.

To analyze \eqref{eq:recursion-coarse}, which involve convolutions, it is convenient to introduce the moment generating functions:
\begin{equation}
\hat{p}(h) := \sum_{\mathcal{M}} e^{h \mathcal{M}} p_{\mathcal{M}} \,,
\end{equation} 
and {$\hat{a}$ and $\hat{b}$} are similarly defined from $a$ and $b$, respectively. Then the recursion relations imply that 
\begin{align} \label{eq:recursion-hat}
&  \hat{p}_{t+1}  = \hat{p}_t^2 +  (\hat{a}'_t)^2 \,,\,  \hat{a}_{t+1}  = 2 \hat{p}_t \hat{a}'_t  \,,\, \hat{b}_{t+1}  =  (\hat{b}'_t)^2 \,,
\end{align}
for any $h$ (its dependence is omitted for brevity). Now, one may readily check that the derivatives of $\hat{p}$ at $h = 0$ are the moments of the outcome distribution, and those of $\hat{a}, \hat{b}$ are the joint moments with $u$ and $v$, respectively. The first moments are as follows:
\begin{align}
&\hat{p}_t(h) = 1 + \left<  \mathcal{M} \right>_t h  + \left<  \mathcal{M}^2 \right>_t h^2 / 2 + \mathcal{O}(h^3), \label{eq:pabexpansion} \\ 
&     \hat{a}_t(h) = \left< \mathcal{M} u \right>_{t} h + \mathcal{O}(h^2) \,,  \hat{b}_t(h) = \left< \mathcal{M} v \right>_{t} h + \mathcal{O}(h^2) \,. \nonumber
\end{align}
Here $\hat{a}_t(0) = \hat{b}_t(0) = 0$ comes from the general property  $\left< \tilde{Q} \right>=\mathbf{1}$ \eqref{eq:Qaverageis1}. Plugging the expansion~\eqref{eq:pabexpansion} back into~\eqref{eq:recursion-hat}, we obtain the recursion relations for the moments. That of the first moment $\left<  \mathcal{M} \right>_{t+1} =2 \left<  \mathcal{M} \right>_{t} $ together with the initial condition~\eqref{eq:init-coarse} implies that  
\begin{equation}
\left<  \mathcal{M} \right>_{t} = 0 \,.
\end{equation}
The first nontrivial moment recursion relations are thus:
\begin{align}
&    \left<  \mathcal{M}^2 \right>_{t+1} = 2 \left<  \mathcal{M}^2 \right>_{t} +  2 \left< \mathcal{M} u' \right>^2_{t}  \label{eq:M2-recursion} \\ 
&  \left< \mathcal{M} u \right>_{t + 1} = 2 \left< \mathcal{M} u' \right>_{t} \,,\, \left< \mathcal{M} v \right>_{t + 1} = 0 \,. \label{eq:Mu-recursion}
\end{align}
Hence, we may lose $\left< \mathcal{M} v \right> $ altogether and replace $  \left< \mathcal{M} u' \right> = \cos (J \pi / 2) \left< \mathcal{M} u \right> $. Then \eqref{eq:Mu-recursion} implies
\begin{equation}\label{eq:Mut}
\left< \mathcal{M} u \right>_t \simeq c^{k-1}  (2 c)^{t}  \,,\, c \equiv \cos (J \pi / 2) \,.
\end{equation}
Now, comparing the two terms in \eqref{eq:M2-recursion}, we see that the encoding-intermediate critical point $J_c=1/2$ is a threshold. When $J > J_c$, we can ignore the second term since the growth generated by the first one ($\sim 2^t$) is much larger:
\begin{equation} \label{eq:M2t-encoding}
\left<  \mathcal{M}^2 \right>_{t} \simeq C 2^t \gg \left< \mathcal{M} u \right>^2_{t}  \quad (J > J_c) \,,
\end{equation}
[the prefactor $C = (1 + c^{2k}) / (1 - 2c^2) $ can be determined by solving explicitly \eqref{eq:M2-recursion}, and the same below].  Meanwhile, when $J < J_c$, the opposite happens:
\begin{equation} \label{eq:M2t-inter}
\left<  \mathcal{M}^2 \right>_{t}  \simeq C'  (2 \cos (J \pi / 2))^{2t} \sim   \left< \mathcal{M} u \right>_t^2  \, (J < J_c), %
\end{equation}
where $C' = c^{2k} / (2c^2 - 1)$. The asymptotic growth rate of the variance $ \left<  \mathcal{M}^2 \right>_{t} $ is non-analytical at $J_c = \frac12$, so the latter is a critical point for the total spin measurement as well. Observe that both pre-factors above diverge at $J = J_c$, where there is a correction to scaling $ \left<  \mathcal{M}^2 \right>_{t} \sim t 2^t $. 
So, the total-spin measurement gives rise to the same encoding-intermediate transition. Since the QD phase is absent, we have an encoding phase at $J > J_c$ and an intermediate phase at $J < J_c$.

We now turn to characterize the two phases in terms of the total-spin measurement. First, when $J > J_c$, we claim that the total spin distribution tends to a Gaussian. To be precise, we will consider the re-scaled total spin  
\begin{equation}
m := 2^{-t/2} \mathcal{M}  \,,
\end{equation}
which by \eqref{eq:M2t-encoding} has a finite variance as $t \to\infty$ ($m$ should not be confused with the earlier notation for a general measurement outcome). The re-rescaling amounts to considering the moment generating function at small $h$, 
\begin{equation}
h =  2^{-t/2} \tilde{h}  \,. 
\end{equation}
Now, the low-moment calculation above applies precisely to the small-$h$ regime, and tells us that $\hat{a}_t, \hat{b}_t \ll \hat{p}_t$ and can be neglected as $t\to\infty$. Hence, the recursion relation~\eqref{eq:recursion-hat} can be approximated by $ \hat{p}_{t+1}  \approx \hat{p}_t^2 $, and the approximation becomes asymptotically exact. Hence, we can fix $t_0 \gg 1 $ and write:
\begin{equation}\label{eq:phat-CLT}
\hat{p}_t(h ) \approx \hat{p}_{t_0}(h)^{2^{t-t_0}} \to e^{c \tilde{h}^2 / 2} \,,\, t \to \infty
\end{equation}
where $c$ is the coefficient in the expansion $\hat{p}_{t_0}(h) = 1 + c \tilde{h}^2 / 2 + \dots$. Eq.~\eqref{eq:phat-CLT} is equivalent to saying that $m$ tends to a centered Gaussian. Its variance is fixed by \eqref{eq:M2t-encoding} above, and depends on $k$. This ``central limit theorem'' for the encoding phase can be also understood as follows: for any $c > 0$,
\begin{equation}
\hat{p}_* = e^{c\tilde{h}^2 / 2} \,,\,  \hat{a}_* =  \hat{b}_*  = 0 
\end{equation} 
is a stable fixed point of the recursion map \eqref{eq:recursion-hat} plus the rescaling $\tilde{h}\mapsto \tilde{h} / {\sqrt{2}}$. 

\begin{figure}
\centering
\includegraphics[width=\columnwidth]{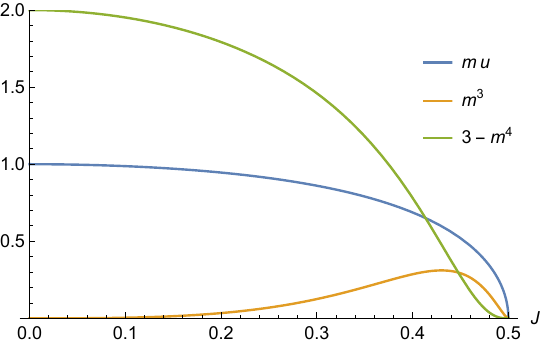}
\caption{The covariance $\left< m u \right>_*$~\eqref{eq:mu}, skewness $\left< m^3 \right>_*$~\eqref{eq:m3}, and negative excessive kurtosis~$\left< 3 - m^4 \right>_*$~\eqref{eq:m4} of the fixed point outcome distribution $m$, normalized so that $\left< m^2 \right> = 1$.}
\label{fig:m3m4}
\end{figure}
When $J < J_c$, it is natural to expect that the recursion map \eqref{eq:recursion-hat} plus the rescaling $\tilde{h}\mapsto \tilde{h} / (2 c)$ goes to a unique stable fixed point upon fixing $(\partial_{\tilde{h}}^2 p)_{\tilde{h} \to 0} = 1$, which is equivalent to $\left< m^2 \right> = 1$. The moments of this fixed point distribution can be computed order by order by expanding \eqref{eq:recursion-hat}. For instance, the skewness and excessive kurtosis of $m$, which characterize its non-Gaussianity, are as follows:
\begin{align}
\left< m^3 \right>_* &=  \frac{3 \cos ^{\frac{3}{2}}(2 \theta ) \tan ^3(\theta )}{(2 \cos (\theta )-1) \left(4 \cos ^3(\theta )-1\right)} \,, \label{eq:m3}\\ 
3-\left< m^4 \right>_*  &= \frac{3 \left(2 c^2-1\right)^2 g(c)}{c^7 (2 c-1)^2 (2 c+1) \left(8 c^4-1\right)} \,. \label{eq:m4}
\end{align}
where $g(c)=16 c^{10}-8 c^9+14 c^8-c^7-6 c^6-11 c^5+4 c^4+7 c^3+2 c^2-c-2$. The covariance between $m$ and $u$ is given by
\begin{equation}\label{eq:mu}
\left< m u \right>_* = \sqrt{2- \sec(\theta)^2} \,.
\end{equation} 
see Fig.~\ref{fig:m3m4} for plots. All the above quantities vanish as $J \to 1/2$ where $m$ becomes Gaussian and uncorrelated with $u$. For $0 < J < 1/2$, $m$ is positively skewed, has negative excessive kurtosis (due to the two peaks), and is positively correlated with $u$. All of this is qualitatively consistent with the numerical results (Fig.~\ref{fig:PMs}) above. 

\subsection{A Clifford model with a direct transition}\label{sec:clifford-tree}
As a last point of this Section, we briefly study a random Clifford analog of the above models, which interpolates between the same $J = 0$ and $J=1$ limits. Curiously, it has a distinct phase diagram, with a direct first order transition from QD to encoding phases.  

\begin{figure}
\centering
\includegraphics[width=.49\columnwidth]{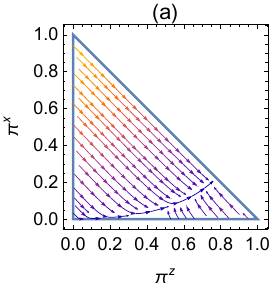}  \includegraphics[width=.49\columnwidth]{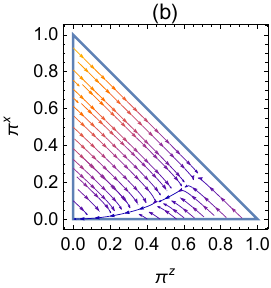}
\caption{The flow generated by the recursion relation~\eqref{eq:recursion-Clifford} of the Clifford model, at $J = 0.45 < J_c$ (a) and $J = 0.55 > J_c$ (b). The asymptotic limit undergoes a discontinuous transition from a QD fixed point to the encoding fixed point. }
\label{fig:Cflow}
\end{figure}
To define the model, it suffices to slightly modify the definition of the isometry in Section~\ref{sec:isometrydef}, by letting the rotation angles $\theta$ be random (and independent): $\theta = 0$ with probability $1-J$, and $\theta = \pi / 2$ with probability $J$. In terms of the parametrization $\tilde{Q} = \mathbf{1} + u \sigma^z + v \sigma^x$~\eqref{eq:defparamuv} and $U^\dagger \tilde{Q} U = \mathbf{1} + u' \sigma^z + v' \sigma^x $ ($U = e^{-i\sigma^y}$ is the rotation), we have
\begin{equation}
(u',v') = \begin{cases}
	(u,v) & \text{with probability $1-J$} \\ 
	(-v, u) & \text{with probability $J$} \,.
\end{cases}
\end{equation}
This model coincides with the deterministic model above at $J = 0$ and $J=1$, but is now Clifford for any $J \in (0,1)$. Then the general argument in Section~\ref{sec:Clifford-gen} implies that the density matrix distribution is supported on the finite set $\{(0,0), (\pm 1, 0), (0, \pm 1)\}$ in terms of $(u,v)$. Also, since $\langle u \rangle = \langle v \rangle = 0$, the distribution depends only on two parameters $\pi^z$ and $\pi^x$ satisfying  $\pi^z \ge 0, \pi^x \ge 0, (1-\pi^z - \pi^z) \ge 0$, as follows,
\begin{align}
\mathbb{P}(\tilde{Q}) & = \frac{\pi^z}2 \left[ \delta(u-1) +   \delta(u+1) \right] \delta(v)\nonumber \\ &+ \frac{\pi^x}2 \left[ \delta(v+1) +  \delta(v-1)  \right]  \delta(u) \nonumber \\  &+ (1-\pi^z - \pi^x) \delta(u)\delta(v) \,.
\end{align}
Using the method of Section~\ref{sec:recursion-uv}, it is not hard to derive a recursion relation for $\pi^z_t$ and $\pi^x_t$ (one can find the same recursion relations using the method of Ref.~\cite{FC}):
\begin{align}
&\pi^z_{t+1} =2  (\pi^z_t (1-J) + \pi^x_t J) - (\pi^z_t (1-J) + \pi^x_t J)^2  \nonumber \\ 
&\pi^x_{t+1} =  (\pi^x_t (1-J) + \pi^z_t J)^2 \,. \label{eq:recursion-Clifford}
\end{align}
The recursion flow, plotted in Fig.~\ref{fig:Cflow}, has a remarkable feature: the $t\to\infty$ limit has a direct discontinuous transition at $J = J_c = 1/2$ from a QD fixed point  ($\pi^z + \pi^x = 1$, $J<1/2$) to the encoding fixed point ($\pi^z = \pi^x = 0$, $J>1/2$). In fact, one can show that \eqref{eq:recursion-Clifford} admits no other fixed points, unless $J = J_c$. At that point, \eqref{eq:recursion-Clifford} has a line of fixed points: $(\pi^z, \pi^x) = (a - a^2 / 4, a^2/4), 0 \le a \le 1$, connecting the perfectly encoding distribution ($a = 0$) to a perfectly QD one ($a=1$). This behavior, which is the origin of the discontinuous transition, is non-generic among Clifford models. Indeed, the model of Ref.~\cite{FC}, which involves more one-body Clifford gates, has a mixed phase and two continuous transitions. 

To conclude, the solution of the above Clifford toy model shows that a direct QD-encoding transition is in principle possible. It will be interesting to find non-Clifford model with a direct QD-encoding transition.  

\section{Conclusion}\label{sec:discussion}
We studied phases of information propagation and the emergence of classical objectivity in a structured environment. We proposed a general framework and a quantitative probe of the different phases: Quantum Darwinism (QD), intermediate, and encoding. We partially solved two similar mean-field models, which exhibit the three phases separated by two continuous transitions of distinct nature. 

The encoding-intermediate transition marks the onset of broadcast, at which point the injected information becomes partially accessible in small fractions of the environment. It can be probed by measuring an extensive (``coarse-grained'') quantity: its non-Gaussian fluctuation (reminiscent of symmetry-breaking) and correlation with the injected information are signatures of the intermediate phase. Such measurements should be in principle accessible in an experimental realization of our mean-field models, and arguably in their non-hierarchical cousins as well.  

In contrast, distinguishing the QD and intermediate phases is more laborious, and requires a fine-grained observation of the fraction. Conceptually this is due to the fact that the QD-intermediate transition breaks a more abstract replica symmetry. Indeed, a replicated perfectly QD distribution (the results of this work are exclusively about the physical $n\to1$ limit)
\begin{align}
&  \sum_m p_m^n \tilde{Q}_m^{\otimes n} \propto \sum_n p_m^n | \varphi_m \rangle^{\otimes n} \langle  \varphi_m  |^{\otimes n}  \nonumber 
\\ &   \in \left( \mathbb{C}^{q} \right)^{n} \otimes \left( \mathbb{C}^{q} \right)^{n} 
\end{align}
is symmetric under the $\mathcal{S}_n \times \mathcal{S}_n$ action that permutes the bras and the kets. By contrast, a replicated encoding/intermediate distribution breaks this symmetry, and favors one particular pairing between the bras and kets~\cite{MIPTrev}. In an experimental realization of our tree model, one may proceed in a hybrid fashion: one measures $F$, computes (classically) the posterior density matrix $\propto \tilde{Q}_m$ by hand from the outcome, and verifies it by measuring $R$ in the direction specified by $\tilde{Q}_m$. In an ideal experiment, one may predict the outcome of the $R$-measure perfectly in the QD phase, and imperfectly in the intermediate phase. Now, the middle classical step would become computationally hard in a non-hierarchical model, due to the operator growth involved in the Heisenberg evolution $Q_m = V^\dagger \pi_m Q$. (A similar difficulty, known as the ``post-selection problem'', is known in the context of measurement-induced transitions~\cite{MIPTnature,li2022cross,garratt2023probing,turkeshi}.)

Despite the potential technical challenges, it is important to note that going beyond hierarchical models may cure the artefacts of the latter, in at least two ways. First, the operator growth can turn a local spin operator into a sum of such terms (by diffusion), and thus accessing effectively a larger environment fraction. Second, tree models have pathological space-time domain walls. Indeed, the two sub-trees of the root do not interact anymore beyond the initial branching. Thus, in our model, with probability $\sim J^2$ (for $J$ small), we may measure large and opposite total spins in the two sub-trees: such discordant amplification \textit{in fine} leads to the absence of the QD phase with coarse-grained measurements. In a non-hierarchical model, such a space-time configuration would have a domain wall cost that grows with the system size $\propto t$, and thus parametrically suppressed. Based on these considerations, we speculate that probing QD-intermediate {transitions} beyond mean-field may require fewer measurements, which leads to a simpler post-selection problem. It is also possible that the enhancement of QD in non-hierarchical systems could lead to a direct QD-encoding transition, bypassing the intermediate phase. 

We therefore advocate that future work on QD-encoding transitions should shift focus onto finite-dimensional and potentially more realistic systems, for example, expanding quantum circuits (with loops), or central spin models~\cite{vidal-mera,leggett-review,LeHur-spinboson}.  From a statistical physics perspective, a natural problem is to characterize the universality class of both transitions. We may attempt to understand the encoding-intermediate transition as a dynamical criticality associated with symmetry breaking~\cite{dynamiccritical,dyncrit-driven}, whereas it will be meaningful to compare the QD-intermediate transition to an entanglement phase transition. Finally, it may be even more important to relate our technical results to the conceptual questions that motivated them at the first place, for example, which phases of quantum information underlie the wave-function collapse as perceived by an agent in the laboratory.

\begin{acknowledgements}
X.C. acknowledges support from a France Berkeley Fund grant  (Project \#24\_2023) and thanks the Laboratoire de Physique Th\'eorique et Mod\`eles Statistiques for hospitality.
\end{acknowledgements}

\bibliography{refs}

\end{document}